\documentclass[
reprint,
nofootinbib,
 amsmath,amssymb,
 aps,
prd,
]{revtex4-1}
\usepackage{lipsum}
\usepackage{dcolumn}
\usepackage{bm}
\usepackage{color}
\usepackage{graphicx,amsmath,amsmath,amssymb}

\def\sfield{\mathcal{S}}
\def\tfield{\mathcal{T}}

\newcommand{\eqnsplit}[1]{\begin{align}\begin{split}#1\end{split}\end{align}}

\definecolor{awesome}{rgb}{1.0, 0.13, 0.32}
\definecolor{azure(colorwheel)}{rgb}{0.0, 0.5, 1.0}

\usepackage{hyperref}
\hypersetup{
    bookmarks=true,         
    unicode=false,          
    pdftoolbar=true,        
    pdfmenubar=true,        
    pdffitwindow=false,     
    pdfstartview={FitH},    
    pdftitle={My title},    
    pdfauthor={Author},     
    pdfsubject={Subject},   
    pdfcreator={Creator},   
    pdfproducer={Producer}, 
    pdfkeywords={keyword1, key2, key3}, 
    pdfnewwindow=true,      
    colorlinks=true,       
    linkcolor=azure(colorwheel),          
    citecolor=awesome,        
    filecolor=azure(colorwheel),      
    urlcolor=awesome           
}

\begin{document}
\title{Quantum tunnelling from vacuum in multidimensions}

\author{Ibrahim Akal}
\email{ibrahim.akal@desy.de}
\affiliation{Theory Group, Deutsches Elektronen-Synchrotron DESY,\\ Notkestra{\ss}e 85, 22603 Hamburg, Germany}

\author{Gudrid Moortgat-Pick}
\email{gudrid.moortgat-pick@desy.de}
\affiliation{II. Institute for Theoretical Physics, University Hamburg, Luruper Chaussee 149, 22761 Hamburg, Germany}

\date{\today}
\begin{abstract}
  The tunnelling of virtual matter-antimatter pairs from the quantum vacuum in multidimensions is studied.
  We consider electric backgrounds as a linear combination of a spatial Sauter field and, interchangeably, certain weaker time dependent fields without poles in the complex plane such as the sinusoidal 
  and Gaussian cases. Based on recent geometric considerations within the worldline formalism,
  we employ the relevant critical points in order to analytically estimate a characteristic threshold for the temporal inhomogeneity.
  We set appropriate initial conditions and apply additional symmetry constraints in order to determine the classical periodic paths in spacetime. Using these worldline instantons, we compute the 
  corresponding leading order exponential factors showing large dynamical enhancement in general.
  We work out the main differences caused by the analytic structure of such composite backgrounds and also discuss the case with a strong temporal variation of Sauter-type.
  \end{abstract}
\maketitle
\section{Introduction}
Virtual matter-antimatter pairs can tunnel from the quantum vacuum in the presence of a static electric background \cite{Heisenberg:1935qt,Schwinger:1951nm,Dittrich:2000zu}.
The field strength $E_\mathrm{S} = m^2$ denotes the critical Schwinger limit for a particle\footnote{The charge has been absorbed into the field strength.} with mass $m$
which is extremely large explaining why this nonperturbative effect could not yet be realised in the laboratory.
However, field strengths of the order $E \sim 10^{-3} E_\mathrm{S}$, and even beyond, 
are expected to be achievable at upcoming strong field facilities.
Recently, there is also a remarkable interest in analogue condensed matter systems
which may lead to a better understanding of this mechanism \cite{Allor:2007ei,Katsnelson:2012cz,Zubkov:2012ht,Fillion-Gourdeau:2015dga,Linder:2015fba,Akal:2016stu,Fillion-Gourdeau:2016izx,Abramchuk:2016afc,Akal:2017vem,PhysRevLett.95.137601}.

Going beyond the static limit the tunnelling rate highly depends on the background structure such that
inhomogeneities in spacetime are capable to trigger a drastic enhancement even far below $E_\mathrm{S}$ \cite{Schutzhold:2008pz,Dunne:2009gi,Bulanov:2010ei,Akal:2014eua,Otto:2015gla,Linder:2015vta,Kohlfurst:2012rb,Hebenstreit:2014lra,Fillion-Gourdeau:2017uss,Akal:2017ilh}.
Notably, in a static magnetic background the process cannot occur \cite{Kim:2000un,Kim:2003qp,Dunne:2004nc},
since the underlying divergent, but alternating expansion of the one-loop Euler-Heisenberg (EH) effective Lagrangian is Borel summable and has no imaginary part \cite{Dunne:2004nc}. In electromagnetic plane wave backgrounds the latter vanishes simply due to symmetry reasons \cite{Schwinger:1951nm}.
In most cases, preferably for purely
electric backgrounds, studies rely on numerical techniques \cite{Ruf:2009zz,Hebenstreit:2010vz,Hebenstreit:2011wk}.
Analytic results have been so far obtained only for certain special cases, see e.g. \cite{PhysRevD.2.1191,Popov:1971iga,Marinov:1977gq,KeskiVakkuri:1996gn,bai2012electron,Dietrich:2003qf,Dunne:2005sx,Kim:2011jw,Strobel:2013vza,Schneider:2014mla,Ilderton:2015qda,Adorno:2016bjx}.
However, a deeper understanding of the impact of more complex backgrounds is highly desirable.
Such setups may particularly be interesting
in the notion of Liouville
integrability \cite{goriely2001integrability} where
constants of motion
would directly be related to the underlying background structure.
However, the latter
may not obey the Maxwell equations in vacuum and 
can lead to complicated symmetries
which makes their identification even more exhausting.

Beyond one-dimensional backgrounds, already purely electric, the identification of the particles is in general highly demanding and hence diagonalising techniques, such as Bogoliubov-like transformations as well as WKB techniques, are difficult to enforce.
In the present work, our focus will be on certain multidimensional ($1+1$) electric backgrounds giving a semiclassical treatment via the worldline formalism in quantum field theory \cite{Strassler:1992zr,Schubert:2001he}.
This approach permits in general a direct multidimensional treatment,
since the imaginary part of the EH effective action
is evaluated on classical periodic paths in spacetime \cite{Affleck:1981bma,Dunne:2005sx,Dunne:2006st,Dumlu:2011cc,bai2012electron,Schneider:2014mla,Ilderton:2014mla,Linder:2015vta,Ilderton:2015qda,Akal:2017ilh}.
Hence, the challenge is to find such so-called worldline instantons, see e.g. \cite{Dietrich:2014ala,Basar:2015xna} for applications in other contexts.

Similar structures arise in trace formulas \cite{Gutzwiller:1971fy,1990JMP31.2952L} relating the state density in systems with
integrable as well as, chaotic, non-integrable classical limit to the properties of periodic orbits
\cite{Aurich:1988yv,SIEBER1990159,RICHENS1981495}
serving as topological equivalences in quantum field theory \cite{muller2004semiclassical,muller2009periodic}. 
Reformulating the problem with the Gutzwiller formula \cite{Dietrich:2007vw} results in a weighted topological sum \cite{berry1976closed} evaluated on grouped orbits
where the fluctuation prefactors can be collected in a single determinant specified by the associated monodromy matrix \cite{MuratoreGinanneschi:2002tm}.
Interestingly, such orbits may become multiple periodic in
spatiotemporal backgrounds \cite{Dumlu:2015paa}.
Assuming that the Hamiltonian defines an integrable dynamical system,
the latter attribute may indicate the existence of an additional constant of motion reflected by continuous, smooth trajectories in the Poincar{\'e} sections \cite{berry1976closed,shivamoggi2014nonlinear} which may provide, together with the Maslov index \cite{Sun2017}, valuable information \cite{Dumlu:2015paa}.
Those aspects clearly reinforce the complications regarding the identification of symmetries governed by such multidimensional systems.

This paper is organised as follows:
in Sec.~\ref{sec:WLIs} we briefly make the connection to the worldline approach.
We discuss general properties of the resulting instanton equations for electric backgrounds composed of a spatial Sauter field and, interchangeably, certain time dependent fields without poles in the complex plane such as the
sinusoidal and Gaussian types. Both spatial as well as temporal fields, we assume to be pointing in the same direction.
Sec.~\ref{sec:strong-S} is the main part of this paper, studying the case where the spatial dependence dominates in strength.
Compared to \cite{Akal:2017ilh}
we extend the reflection picture for the present multidimensional case
and simplify the underlying system of differential equations applying the corresponding critical points. This allows to accomplish certain analytical predictions for the impact of such backgrounds based on nonperturbative computations.
We show that the results substantially differ from the case with two linearly combined Sauter fields \cite{Schneider:2014mla}.
For an appropriate initial value problem, we accomplish numerical techniques via additional symmetry constraints and calculate the corresponding worldline instantons. Using these instantons
we compute the leading order exponential factors.
In Sec.~\ref{sec:strong-T}
we discuss the case of a double Sauter background with a dominant time dependent part. We calculate the corresponding worldline instantons and briefly comment on some essential differences compared to previous findings in \cite{Schneider:2014mla}.
In Sec.~\ref{sec:conc} we finish with a brief conclusion.
App.~\ref{app:eff-strength} includes some extensive discussion on the critical Keldysh parameter.

Throughout this paper we work with natural units $c = 1$ and $\hbar = 1$.
\section{Tunnelling instantons}
\label{sec:WLIs}
\subsection{Stationary points}
The tunnelling probability in the presence of a classical electromagnetic gauge field (background) is 
\begin{align}
\mathcal{P} = 1 - e^{- 2 \Gamma_\mathrm{v}}
\end{align}
where the decay rate $\Gamma_\mathrm{v}$ is determined by the imaginary part of the EH effective action \cite{Schwinger:1951nm,Affleck:1981bma}. 
We use the worldline formalism \cite{Strassler:1992zr,Schubert:2001he} and
focus on the weakly coupled regime neglecting contributions from the dynamical gauge field \cite{Affleck:1981bma}.
Furthermore, we will restrict ourselves on the leading order term in $\Gamma_\mathrm{v}$, the pair creation rate \cite{Cohen:2008wz}, which we denote as $\Gamma$.
The semiclassical result for $\Gamma$ reads
\begin{align}
 \Gamma \simeq e^{-\mathcal{W}_0}
 \label{eq:Gamma-gen}
 \end{align}
where $\mathcal{W}_0$, the stationary worldline action, is obtained after evaluating the action
\begin{align}
 \mathcal{W} = m a + i \oint du\ \dot x \cdot \mathcal{A}(x_\mu),
\end{align}
here for spin zero particles with $\mathcal{A}_\mu$ being the background vector potential and $a$ a constant,
on the periodic Euclidean instanton path, i.e. $x_\mu(0)=x_\mu(1)$, determined by
\begin{align}
 m \ddot x_\mu = i a \mathcal{F}_{\mu \nu} \dot x_\nu.
 \label{eq:instanton-eqs}
\end{align}
Note that the quantum fluctuation prefactor in \eqref{eq:Gamma-gen} is set to unity, see e.g. \cite{Dunne:2006st,Schneider:2016vrl} for detailed studies.
The invariant obeys $a^2 = \dot x^2$ because of the anti-symmetric field tensor $\mathcal{F}_{\mu \nu}$.
We assume the electric background to be oriented in $\hat x_3$ direction
where the spatial part is represented by the scalar potential $\mathcal{A}_4(x_3)$ and the temporal part by a vector potential $\mathcal{A}_3(t)$. Due to simplifying reasons, both parts shall be described by even functions in $t$ and $x_3$, respectively.
After the rotation in the complex plane ($t \rightarrow i x_4$), we get
\eqnsplit{
  \mathcal{A}_3(x_4) = - i \frac{E_\omega}{\omega} \tfield(\omega x_4),\quad
  \mathcal{A}_4(x_3) =  i \frac{E_k}{k} \sfield(k x_3),
  \label{eq:eucl-vecpotential}
}
where $\tfield$ and $\sfield$, consequently, become odd functions in the corresponding spacetime coordinates. Here, $E_k,E_\omega$ denote the field strengths and $k,\omega$ the wavenumber and frequency, respectively.
Inserting the latter expressions into \eqref{eq:instanton-eqs}, we end up with the following system of differential equations
\eqnsplit{
  \ddot x_4 &= + \frac{a E_k}{m} \bigg[ \epsilon \frac{\partial_4 \tfield(\omega x_4)}{\omega} + \frac{\partial_3 \sfield(k x_3)}{k} \bigg] \dot x_3,\\
  \ddot x_3 &= - \frac{a E_k}{m} \bigg[ \epsilon \frac{\partial_4 \tfield(\omega x_4)}{\omega} + \frac{\partial_3 \sfield(k x_3)}{k} \bigg] \dot x_4,\\
  \ddot x_1 &= \ddot x_2 = 0.
  \label{eq:full-instanton-eqs}
}
where $\epsilon := E_\omega / E_k$ has been defined for the sake of convenience.
Next, we specify the spatial part to be a bell-shaped Sauter field described by
\begin{align}
  \sfield_\text{Sauter}(k x_3) = \tanh(k x_3)
  \label{eq:spat-field}
\end{align}
and introduce, due to conventional reasons, the following dimensionless quantities
\eqnsplit{
  \gamma_k = \frac{m k}{E_k},\qquad \gamma_\omega = \frac{m \omega}{E_k},
}
which are usually referred to as the spatial and temporal Keldysh parameter, respectively.

For the temporal dependence we will choose between two different profiles described by
\eqnsplit{
\tfield_\text{sinusoid}(\omega x_4) &= \sinh(\omega x_4),\\
\tfield_\text{Gaussian}(\omega x_4) &= \sqrt{\pi}\mathrm{erfi}(\omega x_4)/2.
\label{eq:temp-fields}
}
A temporal Sauter field, i.e. $\tfield_\text{Sauter}(\omega x_4) = \tan(\omega x_4)$, has been both analytically and numerically investigated in the limit $\epsilon \ll 1$
leading to an enormous enhancement due to instanton reflections at poles in the complex plane \cite{Schneider:2014mla}. 
This enhancement is not restricted to this specific case and is expected to apply in general for any time dependent field with a distinct pole structure in the instanton plane, see e.g. \cite{Akal:2017ilh}.
Now, we extend such considerations for poleless fields as introduced in \eqref{eq:temp-fields}.
Laser fields  have an oscillatory structure leading to substantial interference effects in phase-space \cite{Dumlu:2011rr,Orthaber:2011cm,Akal:2014eua}. 
Such setups motivate investigations for time dependent backgrounds entailing an oscillatory sub-cycle structure.
For the field examples introduced above,
a spatially inhomogeneous field is closely related to a temporal one
by the analytic continuation $\gamma_\omega \rightarrow i \gamma_k$. This correspondence is automatically included in the worldline instanton approach \cite{Dunne:2005sx,Dunne:2006st}.
In the remaining part the dimensional quantities $x_3,\ x_4$ and $a$
will be given in units of $[m/E_k]$.

\subsection{Symmetries}

For backgrounds composed of fields as introduced in \eqref{eq:spat-field} and \eqref{eq:temp-fields} the closed instanton paths preserve (discrete) reflection symmetry\footnote{Let us remind that the static field instanton in the two-dimensional plane is maximally symmetric with $C_\infty$.},
\eqnsplit{
x_3 \rightarrow - x_3,\quad
x_4 \rightarrow - x_4,
\label{eq:sym}
}
i.e. isomorphic with $C_2$.
In this case one can set the starting point on the solution path satisfying, for instance, $x_3(0) \neq 0$ and $x_4(0)=0$.
Afterwards, from \eqref{eq:sym} we get $\dot x_3(0)=0$ and therefore $\dot x_4(0) = a$, which is a direct consequence of the instanton periodicity.
We may conclude
\eqnsplit{
 x_3(0)=x_3(1)=-x_3(1/2) &\neq 0,\\
 x_3(1/4)=x_3(3/4) &=0,\\
 x_4(0)=x_4(1/2)=x_4(1) &=0,\\
 x_4(1/4)=-x_4(1/4) &\neq 0.
 \label{eq:sym-rels}
}
As we will see, such relations lead to useful constraints improving numerical methods in order to find the correct instanton solutions, see Sec.~\ref{subsec:restricts}. 
\section{Strong spatial dependence}
\label{sec:strong-S}

\subsection{Analytical approximations}
\label{subsec:ana-approx}
We suppose a dominant spatial dependence, $\epsilon \ll 1$, such that the terms proportional to $\epsilon$ in \eqref{eq:full-instanton-eqs} can be neglected, except the cases, where the contribution from $\tfield$ counterbalances the smallness of $\epsilon$.
For $\tfield_\mathrm{Sauter}$ this happens at the pole $x_4^\mathrm{ref} = \pi/(2 \gamma_\omega)$ which serves as a reflection point \cite{Schneider:2014mla}.
For the sinusoidal and Gaussian fields the situation is not so obvious.
However, if
the Keldysh parameter of the weak field is much larger than the critical threshold, whose presence is characteristic for the dynamically assisted mechanism \cite{Schutzhold:2008pz,Dunne:2009gi,Schneider:2014mla,Akal:2017ilh}, a
similar criterion applies even for poleless fields \cite{Akal:2017ilh}.
Those effective reflection points apply for backgrounds depending on space and time as in \eqref{eq:eucl-vecpotential} as well.
For $k \rightarrow 0$ we find the largest contribution from $\sfield_\mathrm{Sauter}$. This follows due to its Euclidean structure that remains bounded from above.
Therefore, we fix $k = 0$ and determine the critical point $x_4^\mathrm{ref}$ for which the spatial contribution becomes negligible compared to the weak temporal one. Since in this static limit the maximal contribution from the spatial part is reached, $x_4^\mathrm{ref}$ may also apply for $k > 0$.
The effective reflection points read as
\begin{align}
\begin{split}
 &{\footnotesize{\textbf{sinusoidal}}}\quad x_4^\mathrm{ref}(\gamma_\omega,\epsilon) \approx \frac{\mathrm{arcsinh} (\gamma_\omega / \epsilon)}{\gamma_\omega},\\
 &{\footnotesize{\textbf{Gaussian}}}\quad x_4^\mathrm{ref}(\gamma_\omega,\epsilon) \approx \frac{\sqrt{\ln(1/\epsilon)}(1+\xi)}{\gamma_\omega},
\end{split}
\label{eq:x4refs}
\end{align}
revealing an additional $\epsilon$ dependence which will have interesting consequences for the tunnelling rate.
The detailed derivation of $\xi$ has been accomplished in \cite{Akal:2017ilh}.

Following this reflection picture, one can analytically integrate the approximated instanton equations to get
\eqnsplit{
\dot x_4 &\approx a \frac{\mathcal{S}_\text{Sauter}(k x_3)}{\gamma_k} + a \mathcal{R},\\
\dot x_3 &\approx a \sqrt{ 1 - \left( \frac{\mathcal{S}_\text{Sauter}(k x_3)}{\gamma_k} + \mathcal{R} \right)^2 }
\label{eq:approx-instanton-eqs}
}
after applying the relation $a^2 = \dot x_3^2 + \dot x_4^2$, cf. \cite{Schneider:2014mla}. Here, $\mathcal{R}$ represents a dimensionless reflection constant which determines the velocity $\dot x_4(\pm 1/4)$ where $x_3 = 0$. Using $1/4 = \int_0^{x_3^\mathrm{turn}} dx_3\ \frac{1}{\dot x_3}$,
which is justified due to the underlying instanton symmetry \eqref{eq:sym}, and inserting the second expression of \eqref{eq:approx-instanton-eqs} into the latter integral, the invariant $a$ satisfies
\begin{align}
a \approx 4 \int_0^{x_3^\mathrm{turn}} dx_3\ \frac{1}{\sqrt{ 1 - \left( \frac{\sfield_\text{Sauter}(k x_3)}{\gamma_k} + \mathcal{R} \right)^2 }}
\label{eq:a-cond}
\end{align}
where the upper integration limit, the spatial turning point $x_3^\mathrm{turn}$, is determined by
\begin{align}
\sfield_\text{Sauter}(k x_3^\mathrm{turn}) + \gamma_k \mathcal{R} = \gamma_k.
\label{eq:x3crit-cond}
\end{align}
The constant $\mathcal{R}$ can be computed using $x_4^\mathrm{ref} = \int_0^{x_3^\mathrm{turn}} dx_3\ \frac{\dot x_4}{\dot x_3}$ and inserting \eqref{eq:approx-instanton-eqs} which gives the following implicit condition
\begin{align}
  x_4^\mathrm{ref} \overset{!}{=} \int_0^{x_3^\mathrm{turn}} dx_3\ \frac{ \mathcal{S}_\text{Sauter}(k x_3) + \gamma_k \mathcal{R} }{\sqrt{ \gamma_k^2 - \left( \sfield_\text{Sauter}(k x_3) + \gamma_k \mathcal{R} \right)^2 }}.
  \label{eq:R-cond}
\end{align}
In case of reflection one has $\mathcal{R} \neq 0$. However, setting $\mathcal{R} = 0$ and
replacing $x_4^\mathrm{ref}$ on the LHS by the appropriate critical point\footnote{The critical temporal Keldysh paramter we assume to be determined by the critical point $x_4^\mathrm{crit}$ where both fields start to contribute equally, i.e. $\mathcal{S} = \epsilon \mathcal{T}$.} denoted as $x_4^\mathrm{crit}$ at which the weak field contribution starts to become dominating,
we can straightforwardly compute $\gamma_\omega^\mathrm{crit}$.
Such critical points can be obtained via perturbations around the following intersection points
\begin{align}
\begin{split}
 &{\footnotesize{\textbf{sinusoidal}}}\quad x_4^\mathrm{int}(\gamma_\omega,\epsilon) = \frac{\mathrm{arccosh}(1/\epsilon)}{\gamma_\omega},\\
 &{\footnotesize{\textbf{Gaussian}}}\quad x_4^\mathrm{int}(\gamma_\omega,\epsilon) = \frac{\sqrt{\ln(1/\epsilon)}}{\gamma_\omega}
\end{split}
\label{eq:crit-values}
\end{align}
such that
\begin{align}
x_4^\mathrm{crit} = x_4^\mathrm{int}(1-\Delta)
\label{eq:x4crit}
\end{align}
with the corresponding correction $\Delta$ explicitly computed up to order $\mathcal{O}(\Delta^2)$, see App. B in \cite{Akal:2017ilh}.
For fields with a pole structure both $x_4^\mathrm{ref}$ and $x_4^\mathrm{crit}$ are identical in the limit $\epsilon \ll 1$, i.e. $\Delta = 0$. For the sinusoidal and Gaussian fields which have no poles present, those two points are not equal, i.e. $\Delta > 0$. Generally, the values for the effective reflection points are much larger.
According to \eqref{eq:crit-values}, we find that the threshold depends on $\epsilon$.
This attribute has already been discussed in the purely temporal case \cite{Schneider:2014mla,Linder:2015vta,Akal:2017ilh}, but needs some further modifications for the present spatiotemporal setup, see App.~\ref{app:eff-strength}. Namely, the effective field strength ratio for $\gamma_k > 0$ is
\begin{align}
  \tilde\epsilon = \epsilon \cosh^2\left(\mathrm{arcsinh}\left(\frac{\gamma_k}{\sqrt{1-\gamma_k^2}}\right)\right).
  \label{eq:epsilon-corr}
\end{align}
Here, one may only consider the spatial Sauter field where the RHS follows from $\mathrm{max}\{x_3\}$ determined by the corresponding exact instanton solution, see App.~\ref{app:eff-strength}.
With these modifications and applying the integral result from \cite{Schneider:2014mla}, the critical temporal Keldysh parameter can be generalised to
\begin{align}
 \gamma_\omega^\mathrm{crit} = \gamma_\omega x_4^\mathrm{int} (1-\Delta) \frac{ \gamma_k \sqrt{1-\gamma_k^2}}{\mathrm{arcsin}(\gamma_k)}
 \label{eq:crit-gamma_omega}
\end{align}
with $x_4^\mathrm{int} \equiv x_4^\mathrm{int}(\gamma_\omega,\tilde\epsilon)$, $\Delta \equiv \Delta(\tilde\epsilon)$ and $\tilde\epsilon \equiv \tilde\epsilon(\gamma_k,\epsilon)$.
Inserting the pole for the Sauter field, i.e. $x_4^\mathrm{int} = \pi/(2\gamma_\omega)$ and $\Delta = 0$, leads to the threshold in \cite{Schneider:2014mla}.
\begin{figure}[h]
  \centering
\includegraphics[width=.45\textwidth]{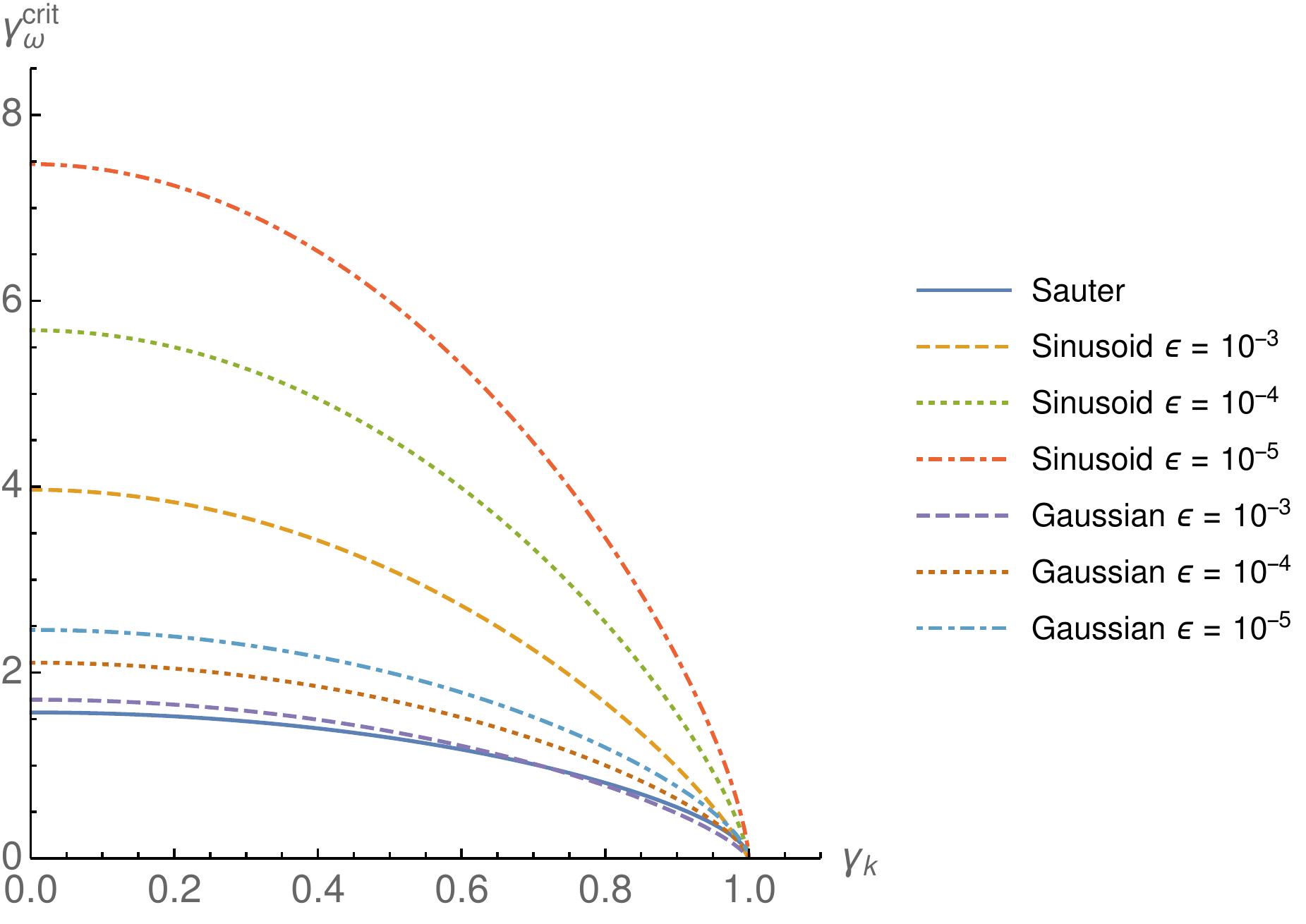}
\caption{Critical temporal Keldysh parameter $\gamma_\omega^\mathrm{crit}$ from \eqref{eq:crit-gamma_omega} plotted versus $\gamma_k$. The values for $\epsilon$ are listed in the plot legend.}
\label{fig:gamma-crit}
\end{figure}
The resulting instantons for all three weak fields are plotted in Fig.~\ref{fig:gamma-crit} where for the poleless cases we have used $\epsilon \in \{10^{-3},10^{-4},10^{-5}\}$. For $\gamma_k = 0$ the critical threshold $\gamma_\omega^\mathrm{crit}$ increases as soon as $\epsilon \rightarrow 0$. In this limit, the Gaussian field leads to a much smaller $\gamma_\omega^\mathrm{crit}$ compared to the sinusoidal case, being in accordance with previous findings in a purely temporal background \cite{Akal:2017ilh}. For $\gamma_k \rightarrow 1$ the threshold for $\epsilon \ll 1$ behaves as $\gamma_\omega^\mathrm{crit} \rightarrow 0$ which reflects delocalisation effects.
For $\gamma_k \geq 1$ the width of the spatial field falls down below the Compton wavelength. Consequently, without additional assistance, i.e. $\gamma_\omega = 0$, the delocalised virtual pair cannot absorb sufficient energy to become a real pair. However, it is expected that the threshold at $\gamma_k = 1$ \cite{Dunne:2005sx,Gies:2005bz,NIKISHOV1970346} will be shifted to larger values for increasing time variations approaching the Compton scale \cite{Gies:2015hia}. This would result in additional energetic multi-photon contributions leading to substantial support.
A similar effect will also apply for $\epsilon \rightarrow 1$, cf. e.g. \cite{Tomaras:2001vs,Ilderton:2014mla,Ilderton:2015qda}.
On the other hand, if $\epsilon > 1$ and $\gamma_\omega \rightarrow 0$, there will be no critical value present for $\gamma_k$.
In this case the tunnelling can entirely be driven by the strong time dependendent term even if the electrostatic energy provided by the weaker spatial term alone is incapable to produce the pair.
More details on this will be discussed in \ref{sec:strong-T}.

Coming back to the present case,
the remaining quantities can be computed according to the following prescription
\begin{align}
  \begin{aligned}
    x_4^\mathrm{ref}\ \&\ \eqref{eq:R-cond}\ \&\ \eqref{eq:x3crit-cond} &\rightsquigarrow \mathcal{R},\\
    \mathcal{R}\ \&\ \eqref{eq:x3crit-cond}\ \&\ \eqref{eq:a-cond} &\rightsquigarrow a,\\
    \mathcal{R}\ \&\ \eqref{eq:x3crit-cond} &\rightsquigarrow x_3^\mathrm{turn}.
  \end{aligned}
  \label{eq:prescripts}
\end{align}
Let us remind that due to restrictions regarding the derivation of $x_4^\mathrm{ref}$ stressed above, see \cite{Akal:2017ilh} for details, we expect the predictions in \eqref{eq:prescripts} to be valid if
\begin{align}
 \gamma_\omega \gg \gamma_\omega^\mathrm{crit}
 \label{eq:expect-cond}
\end{align}
which is the condition for dynamical assistance \cite{Schutzhold:2008pz}.
\subsection{Comparison with numerical results}
\label{subsec:comp-num}
\begin{figure}[h]
  \centering
  \includegraphics[width=.4\textwidth]{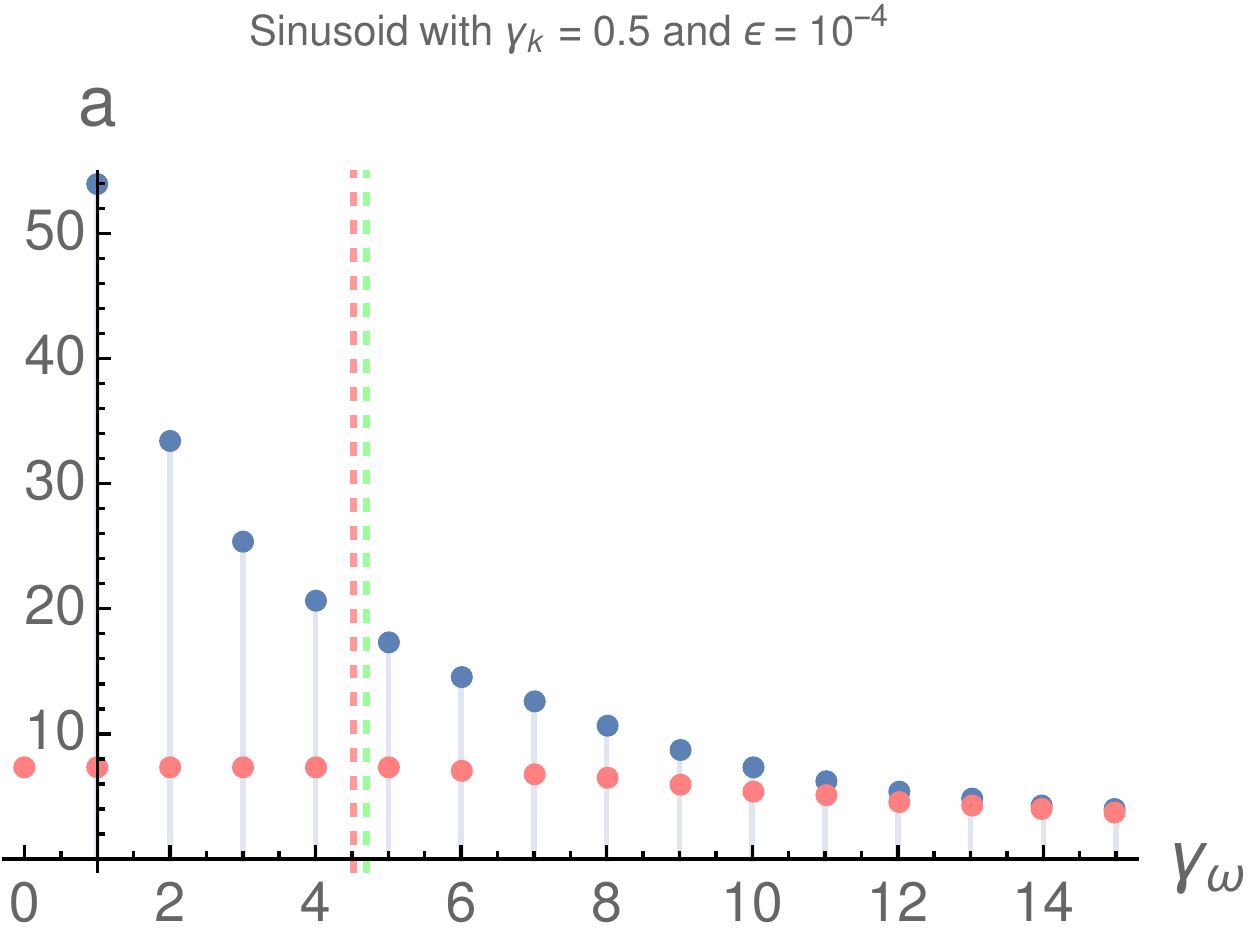}
  \includegraphics[width=.4\textwidth]{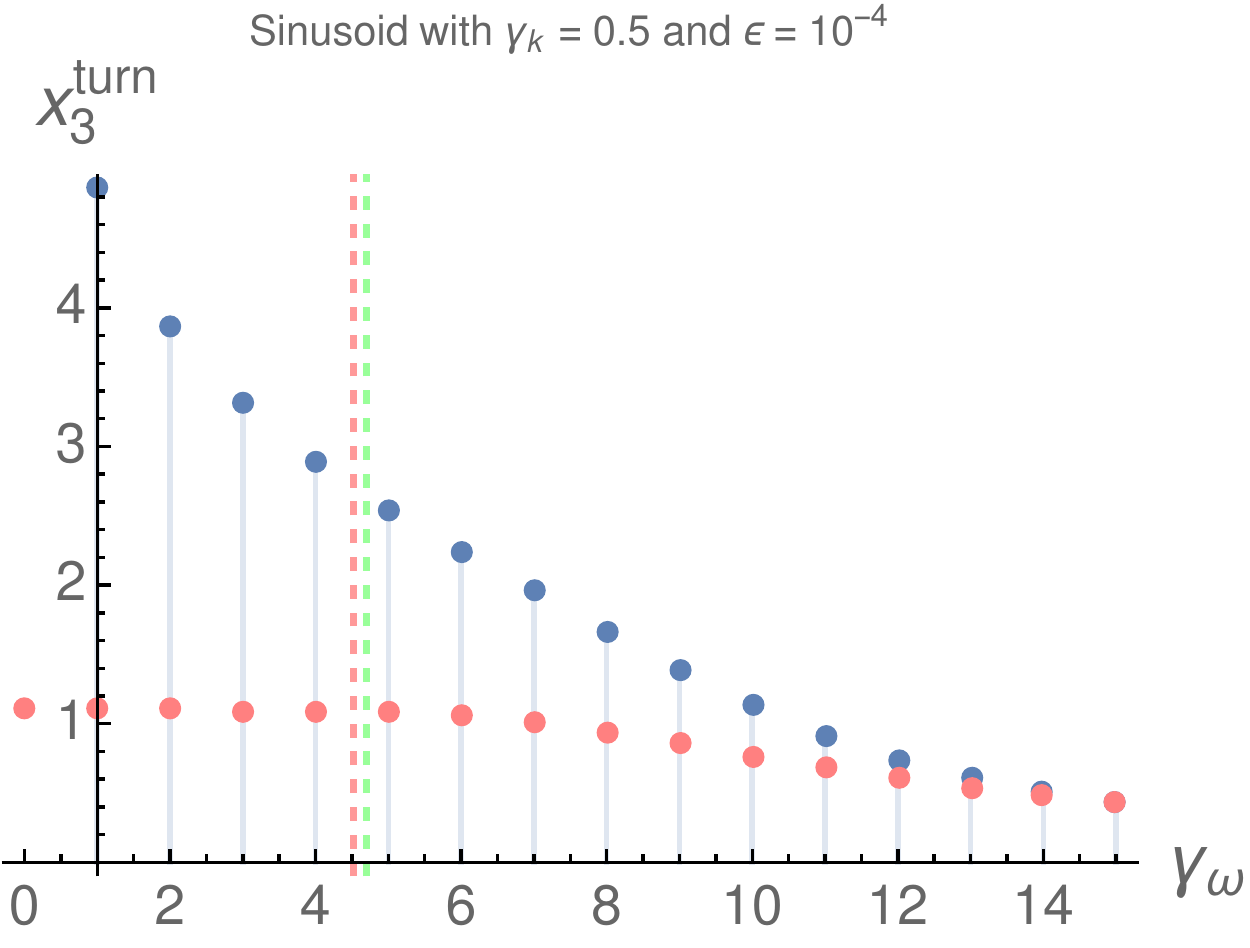}
\caption{$a$ (top) and $x_3^\mathrm{turn}$ (bottom) for time dependent sinusoidal field plotted versus $\gamma_\omega \in \{0,1,\ldots,15\}$: values are computed via numerical shooting (red dots), where starting points $(a^\mathrm{start},x_3^\mathrm{start})$ have been set by hand, and via the prescription in \eqref{eq:prescripts} (blue dots). Remaining field parameters are given as $\gamma_k = 0.5$ and $\epsilon=10^{-4}$. The vertical, dashed, red line is located at $\gamma_\omega^\mathrm{crit}$ from \eqref{eq:crit-gamma_omega}, whereas the dashed, green line has been obtained without replacing $\epsilon$ by the modified parameter $\tilde\epsilon$ from \eqref{eq:epsilon-corr}.}
\label{fig:a-x3turn-sinusoid}
\end{figure}
In this part we will compare the predictions in \eqref{eq:prescripts} with directly obtained numerical results. For solving the system \eqref{eq:instanton-eqs}, having closed periodic paths as solutions, we transform an appropriate boundary value problem
via constructing a convenient multivariate function of an initial condition set by
$a$ and $x_3^\mathrm{turn}$ which we treat via the shooting technique, see also \cite{Dunne:2006ur}. The idea is to reduce the whole task to the problem of finding the root of the multivariate function. This can be easily carried out, for instance, with standard computational tools for which we will first
estimate the required starting point $(a^\mathrm{start},x_3^\mathrm{start})$ by hand.
\begin{figure}[h]
  \centering
  \includegraphics[width=.4\textwidth]{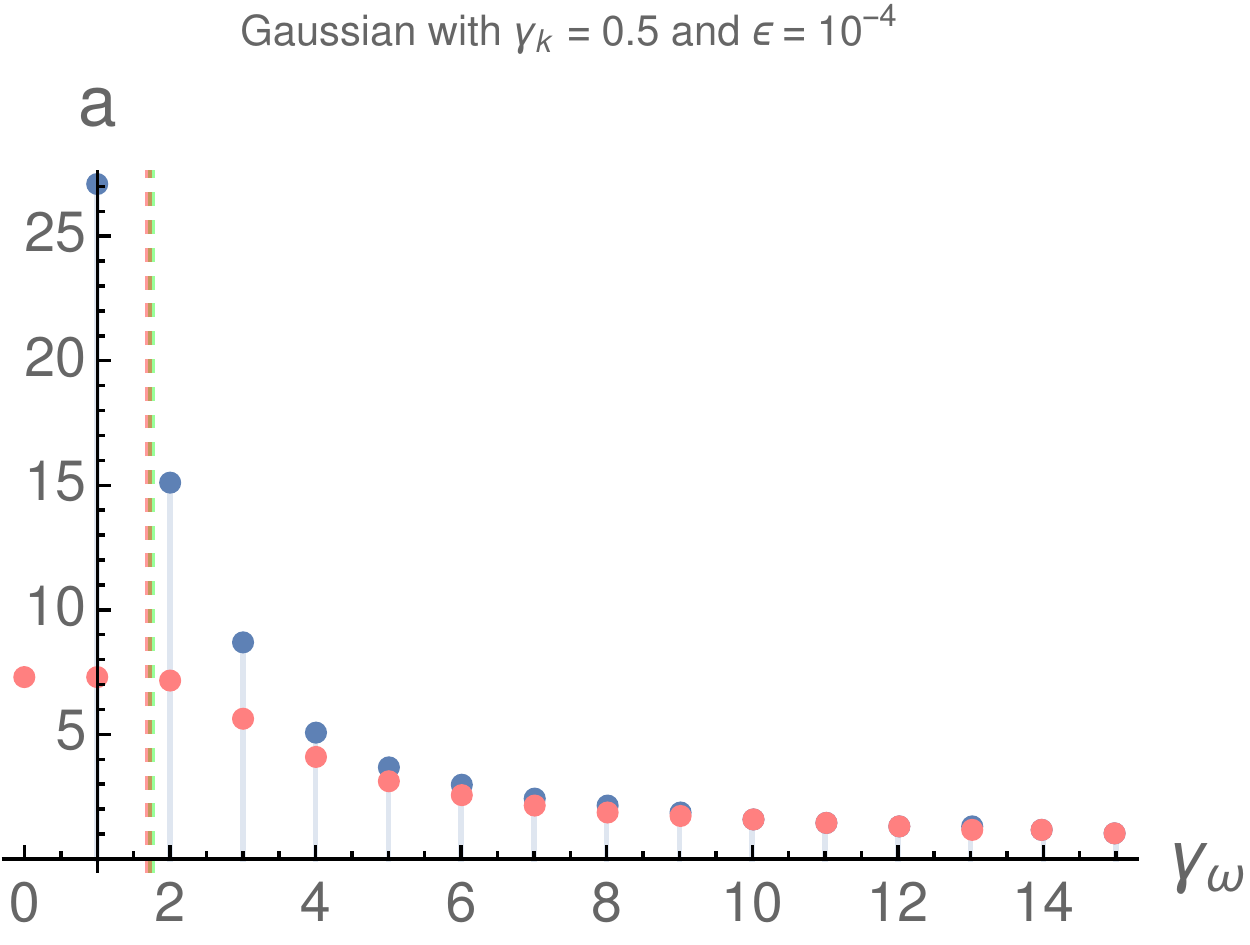}
  \includegraphics[width=.4\textwidth]{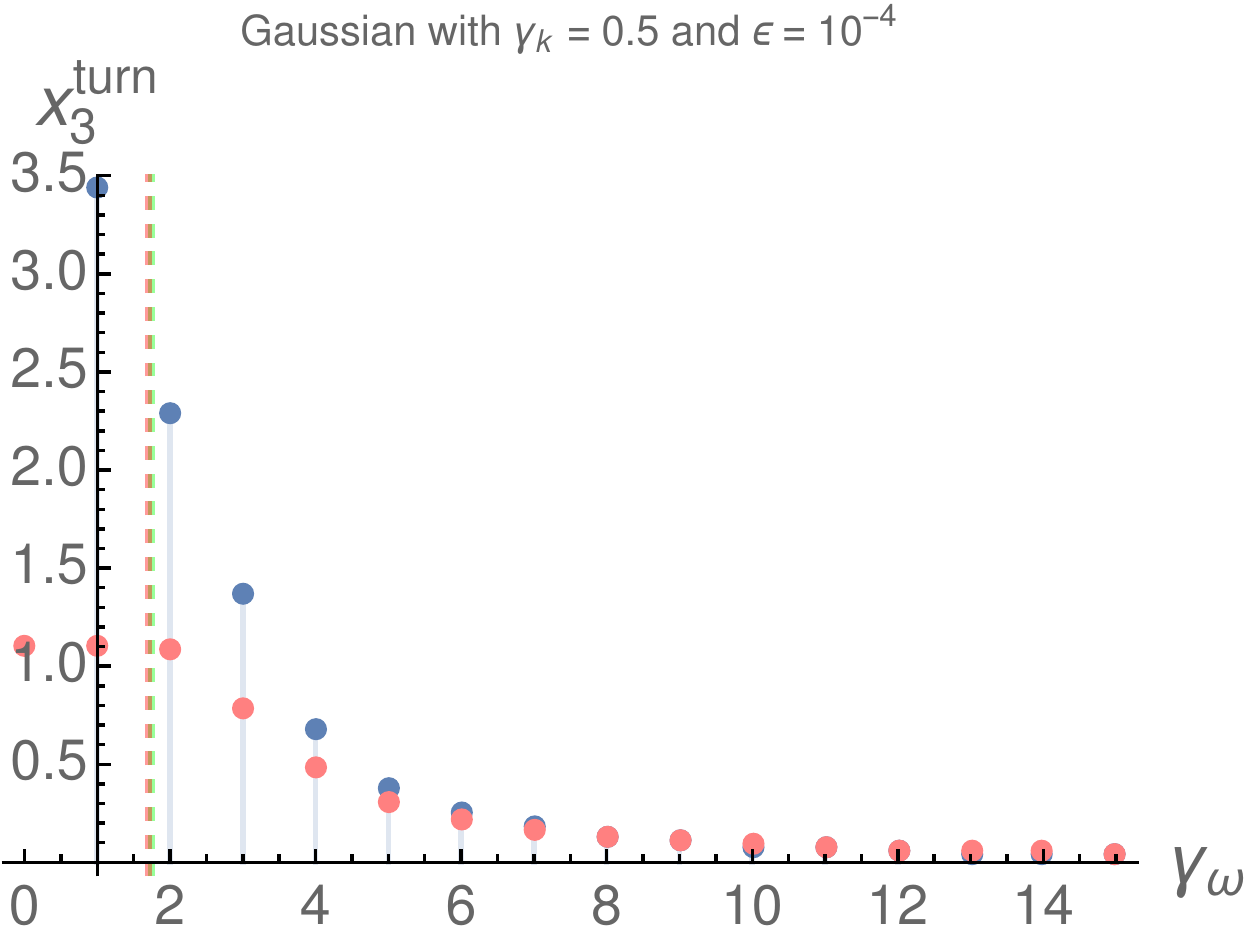}
\caption{$a$ (top) and $x_3^\mathrm{turn}$ (bottom) for time dependent Gaussian field plotted versus $\gamma_\omega \in \{0,1,\ldots,15\}$. Remaining parameters and colors are set as in Fig.~\ref{fig:a-x3turn-sinusoid}.}
\label{fig:a-x3turn-gaussian}
\end{figure}
Taking into account the instanton symmetry \eqref{eq:sym}, we set $\dot x_3(0) = 0$ and $x_4(0) = 0$.
Using the relation for the constant $a$, we end up with the following initial conditions
\eqnsplit{
x_3(0) &= \check x_3,\ \dot x_3(0) = 0,\\
x_4(0) &= 0,\ \dot x_4(0) = \check a.
\label{eq:systems-inis}
}
The function, whose root we have to determine, can then be defined, for instance, as
\eqnsplit{
\pmb{\Omega}_1(\check a,\check x_3) := \left( \begin{array}{c}
x_3(1) - \check x_3\\
x_4(1)
\end{array}
\right) \in \mathbb{R}^2.
}
Note that $\pmb{\Omega}_1$ basically includes only information about the path's periodicity.
Having done this, we can evolute the solution for the pair $(\check a,\check x_3)$ solving the system \eqref{eq:instanton-eqs} with initial conditions \eqref{eq:systems-inis} until a numerical root of $\pmb{\Omega}_1$ is found, which then provides a solution, i.e.
\begin{align}
(a, x_3^\mathrm{turn}) \in \{(\check a,\check x_3)\ |\ \pmb{\Omega}_1(\check a,\check x_3) = 0\}.
\end{align}
The result, however, is very sensitive to the initial starting point which has to be preset for the root finder very carefully. One should note that restricting the solution via $\pmb{\Omega}_1$ may lead to closed paths, but cannot be accepted as a correct solution simply due to violation of \eqref{eq:sym}. Modifications in order to avoid such inconsistencies will be discussed in Sec.~\ref{subsec:restricts}. Nevertheless, apart from those technicalities, we can adjust the starting points for any setting by hand until an appropriate solution is found. Proceeding in this way, the results are depicted in Figs.~\ref{fig:a-x3turn-sinusoid} and \ref{fig:a-x3turn-gaussian}, respectively, fixing the parameters $\gamma_k = 0.5$, $\epsilon = 10^{-4}$ and varying $\gamma_\omega$ as given in the figure captions.
As one can clearly observe, the analytical approximations approach the numerical results for sufficiently large temporal inhomogeneities, i.e. $\gamma_\omega \gg \gamma_\omega^\mathrm{crit}$, being in line with our expectation \eqref{eq:expect-cond}. Furthermore, the critical threshold \eqref{eq:crit-gamma_omega}, both with and without the replacement $\epsilon \rightarrow \tilde\epsilon$, turns out to be remarkably accurate. For  $\gamma_\omega < \gamma_\omega^\mathrm{crit}$ both $a$ and $x_3^\mathrm{turn}$ behave almost constant, reflecting the absence of substantial contributions from the weak field.
\subsection{Starting points and symmetry constraints}
\label{subsec:restricts}
Finding the worldline instantons directly, that is, without tweaking the starting point by hand, requires some refining of the previous strategy. The starting point has to be set accurately in order to find the correct root of $\pmb{\Omega}_1$ which turns out to be very difficult to control.
However, once the correct invariant and spatial turning point is determined, solving the resulting system by incorporating the found root will supply the closed instanton path.
The previous comparisons show that we cannot benefit from \eqref{eq:prescripts} in particular for values in the vicinity of $\gamma_\omega^\mathrm{crit}$.
A possible approach can be pursued as follows:
for $\gamma_\omega \leq \gamma_\omega^\mathrm{crit}$ one simply assesses the starting point as the one that is obtained only for the strong spatial background term, here denoted as $(a_0,x_{0,3}^\mathrm{turn})$, whereas for $\gamma_\omega > \gamma_\omega^\mathrm{crit}$ one decides whether the prediction via \eqref{eq:prescripts} is smaller or larger than $(a_0,x_{0,3}^\mathrm{turn})$. In the former case the analytical approximation can be taken as the corresponding starting point. These steps can be put together as
\eqnsplit{
a^\mathrm{start} &=
\left\{
\begin{array}{ll}
      a_0, & a_0 \leq a\\
      a, & a_0 > a
\end{array}
\right.,\\
x_3^\mathrm{start} &=
\left\{
\begin{array}{ll}
      x_3^{0,\mathrm{turn}}, & x_3^{0,\mathrm{turn}} \leq x_3^\mathrm{turn}\\
      x_3^\mathrm{turn}, & x_3^{0,\mathrm{turn}} > x_3^\mathrm{turn}
\end{array}
\right.
}
where
\eqnsplit{
a_0 = \frac{2 \pi}{\sqrt{1 - \gamma_k^2}},\quad
x_{0,3}^\mathrm{turn} = \frac{1}{\gamma_k} \mathrm{arcsinh}\left( \frac{\gamma_k}{\sqrt{1 - \gamma_k^2}} \right),
}
see App.~\ref{app:eff-strength}.
As a last step, we replace the previous function by
\eqnsplit{
&\pmb{\Omega}_2(\check a,\check x_3)\\
&:= \left( \begin{array}{c}
x_3(1/4) + x_3(3/4) + x_4(1/2) - x_4(1)\\
x_4(1/4) + x_4(3/4) + x_3(1/4)  - x_3(3/4)
\end{array}
\right)
\\
}
for which we have explicitly imposed the constraints from \eqref{eq:sym-rels}.
The solution to the problem is then
\begin{align}
(a, x_3^\mathrm{turn}) \in \{(\check a,\check x_3)\ |\ \pmb{\Omega}_2(\check a,\check x_3) = 0\}.
\end{align}
It turns out that searching for a numerical root of $\pmb{\Omega}_2$ is much more robust and accurate for finding the correct solution which leads to closed paths keeping the instanton symmetry preserved.

\subsection{Worldline instantons}
\label{subsec:WLIs}
\begin{figure}[!h]
  \centering
\includegraphics[width=.17\textwidth]{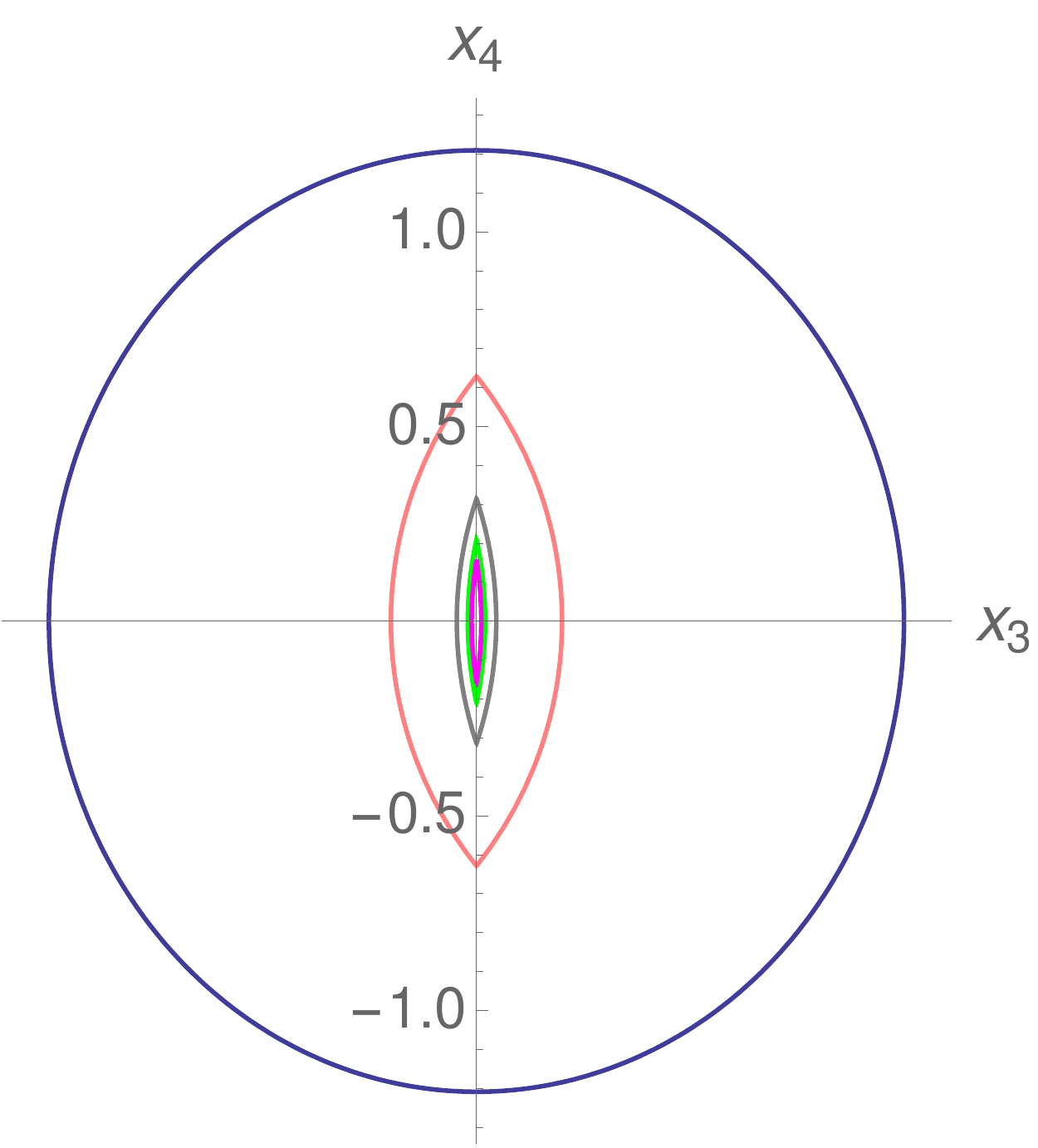}
\includegraphics[width=.30\textwidth]{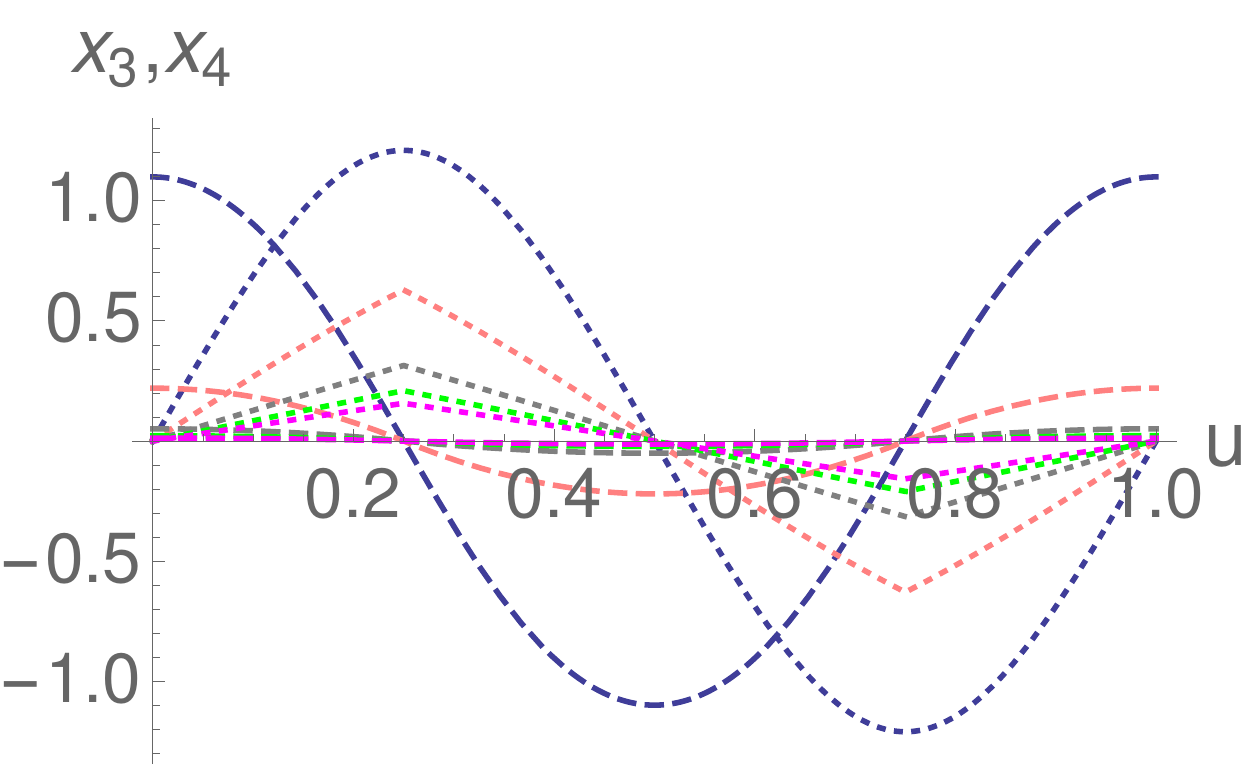}
\caption{
Worldline instantons for superimposed temporal Sauter field with $\gamma_\omega \in \{0.001,2.5,5,7.5,10\}$ (from blue, outer path, to magenta, inner path). In the right panel the components $x_3$ (dashed) and $x_4$ (dotted) are separately plotted. Remaining field parameters are chosen as $\gamma_k = 0.5$ and $\epsilon = 10^{-4}$.
}
\label{fig:WLIs-sauter}
\end{figure}

\begin{figure}[!h]
  \centering
\includegraphics[width=.17\textwidth]{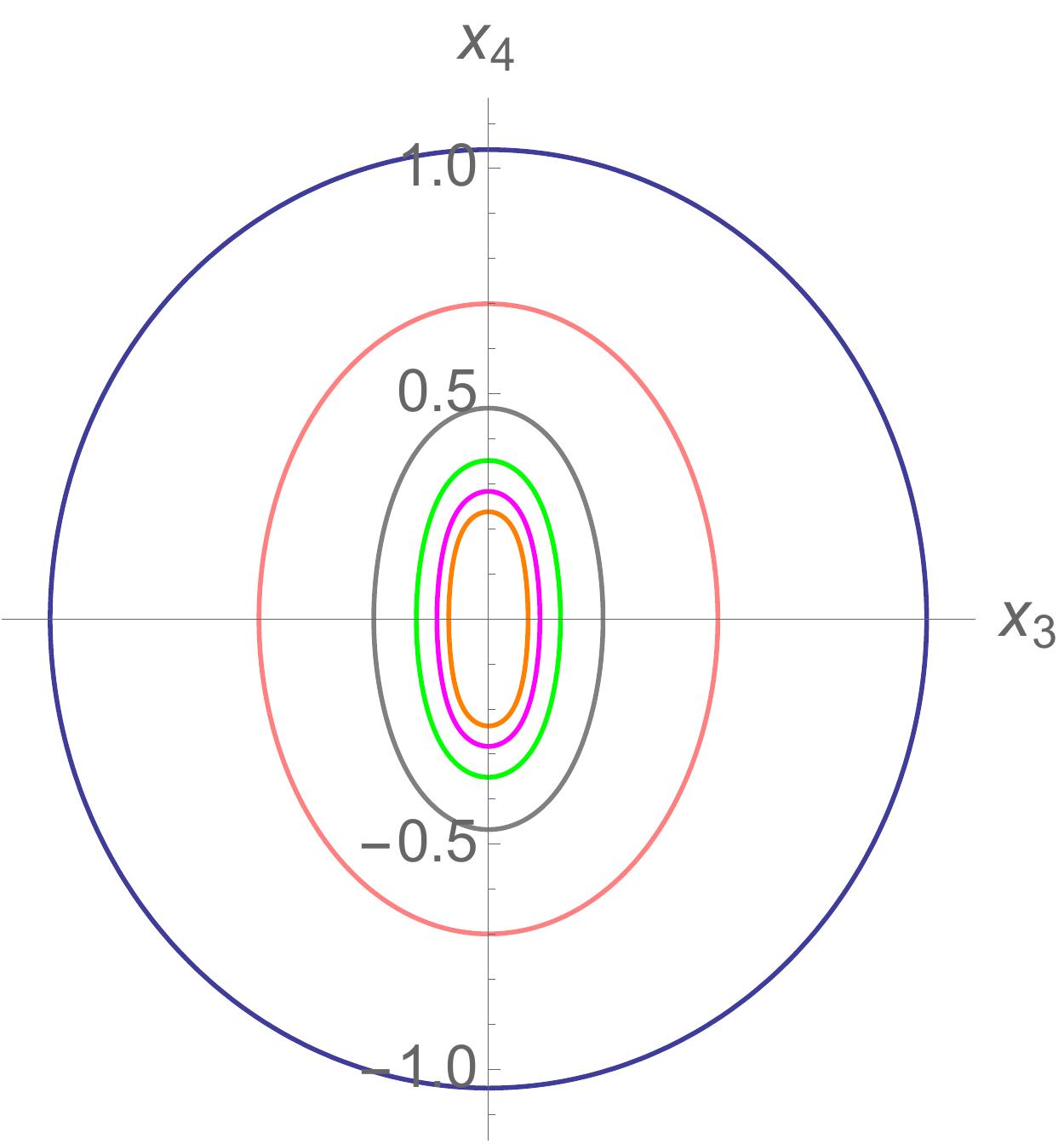}
\includegraphics[width=.30\textwidth]{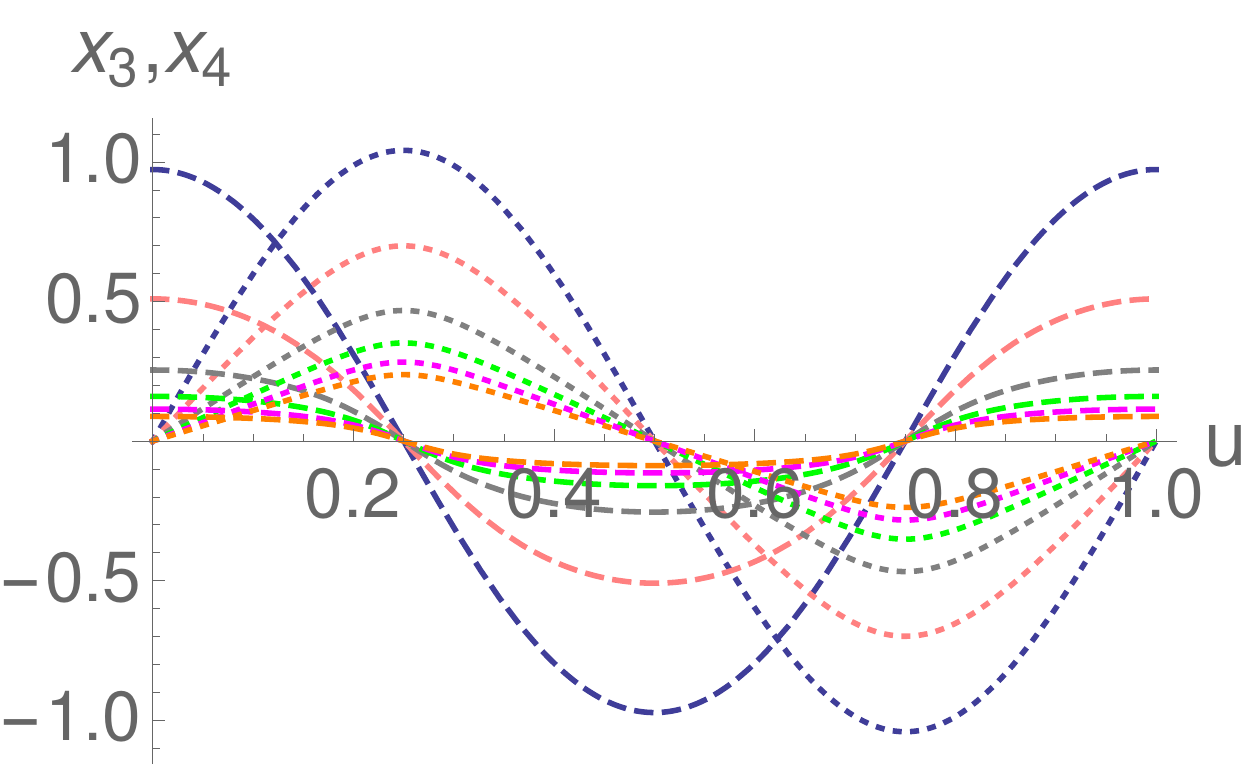}
\includegraphics[width=.17\textwidth]{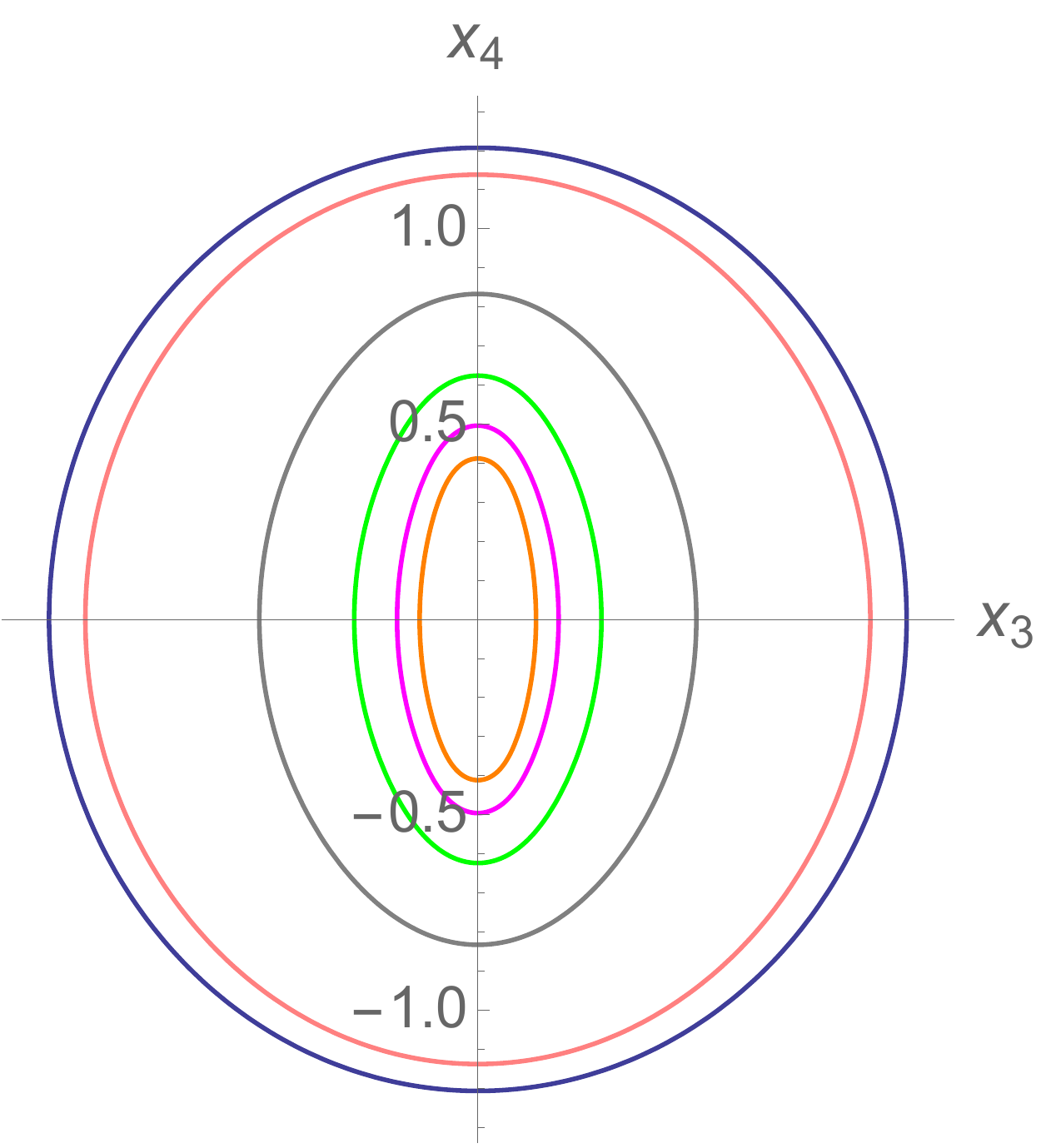}
\includegraphics[width=.30\textwidth]{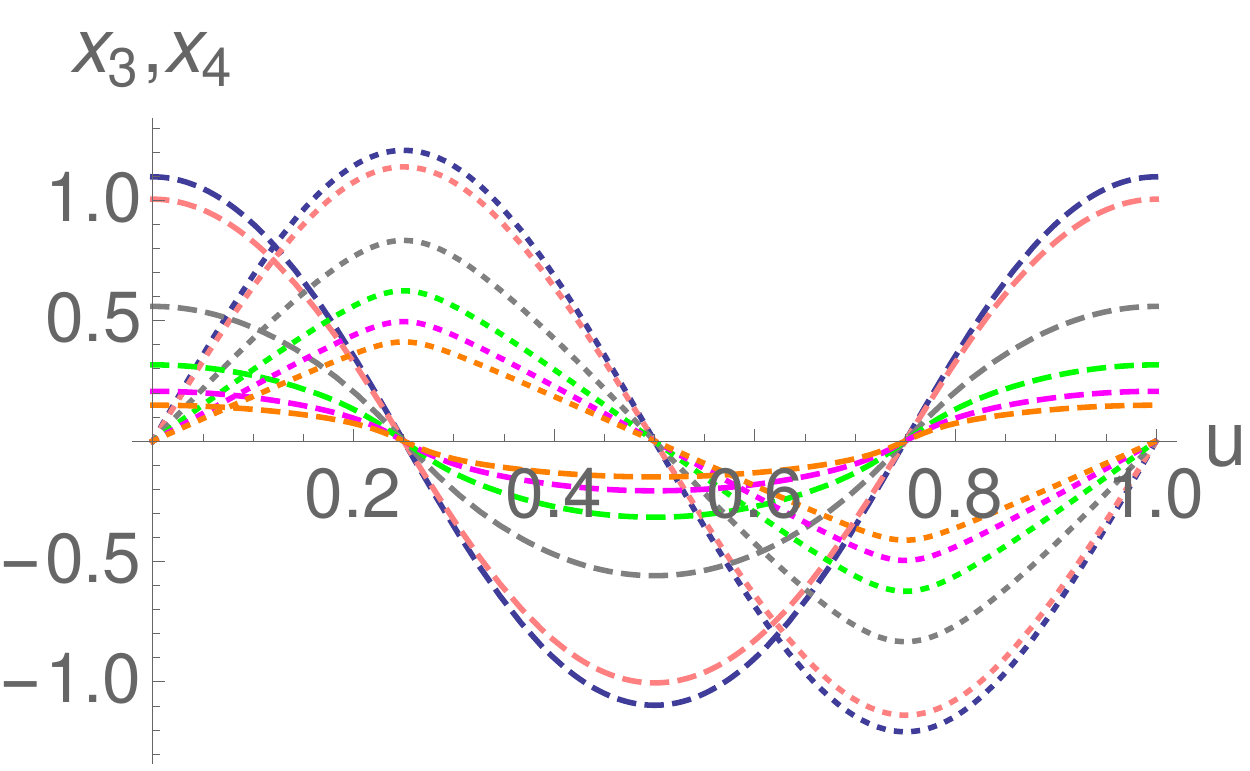}
\includegraphics[width=.17\textwidth]{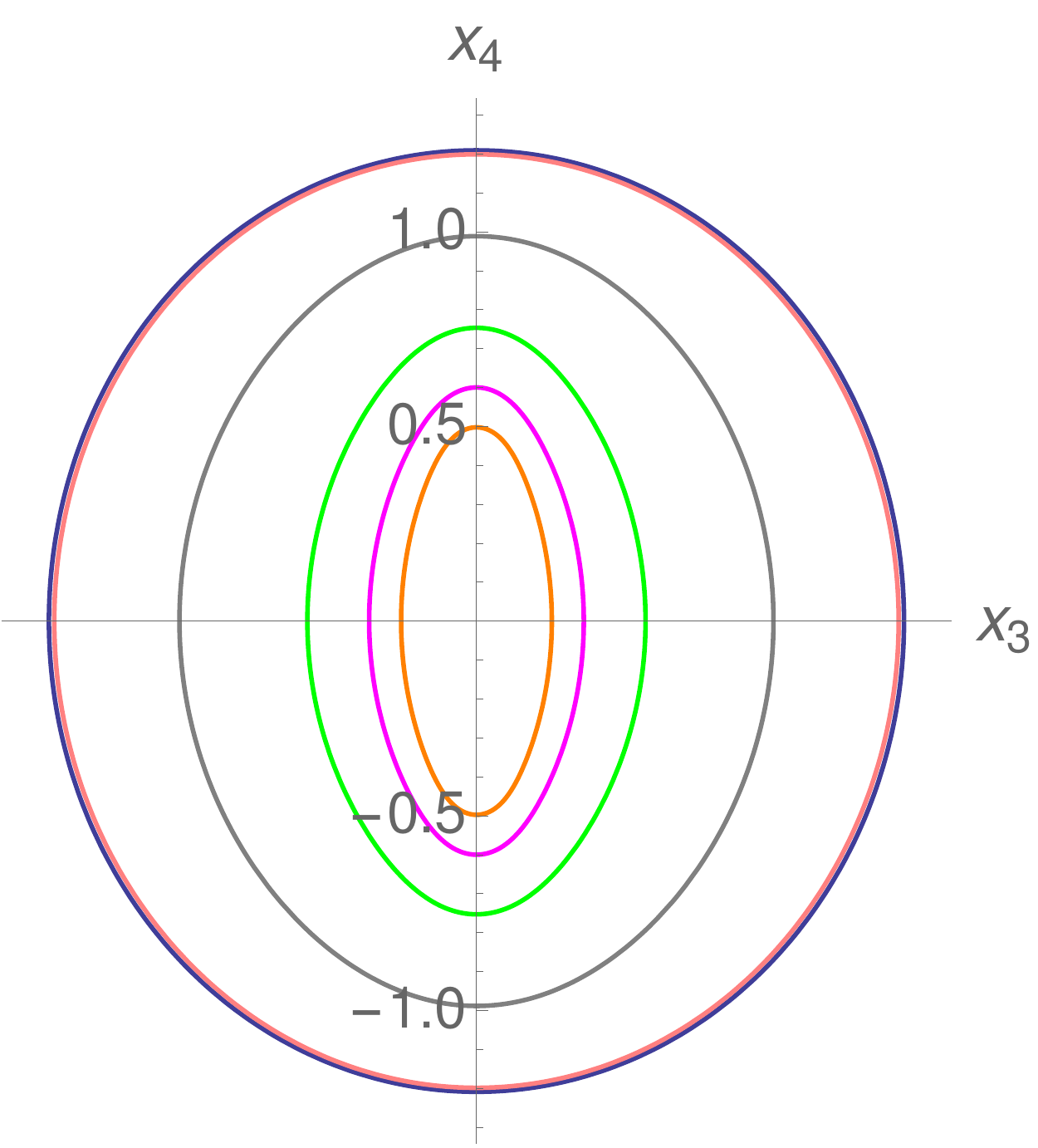}
\includegraphics[width=.30\textwidth]{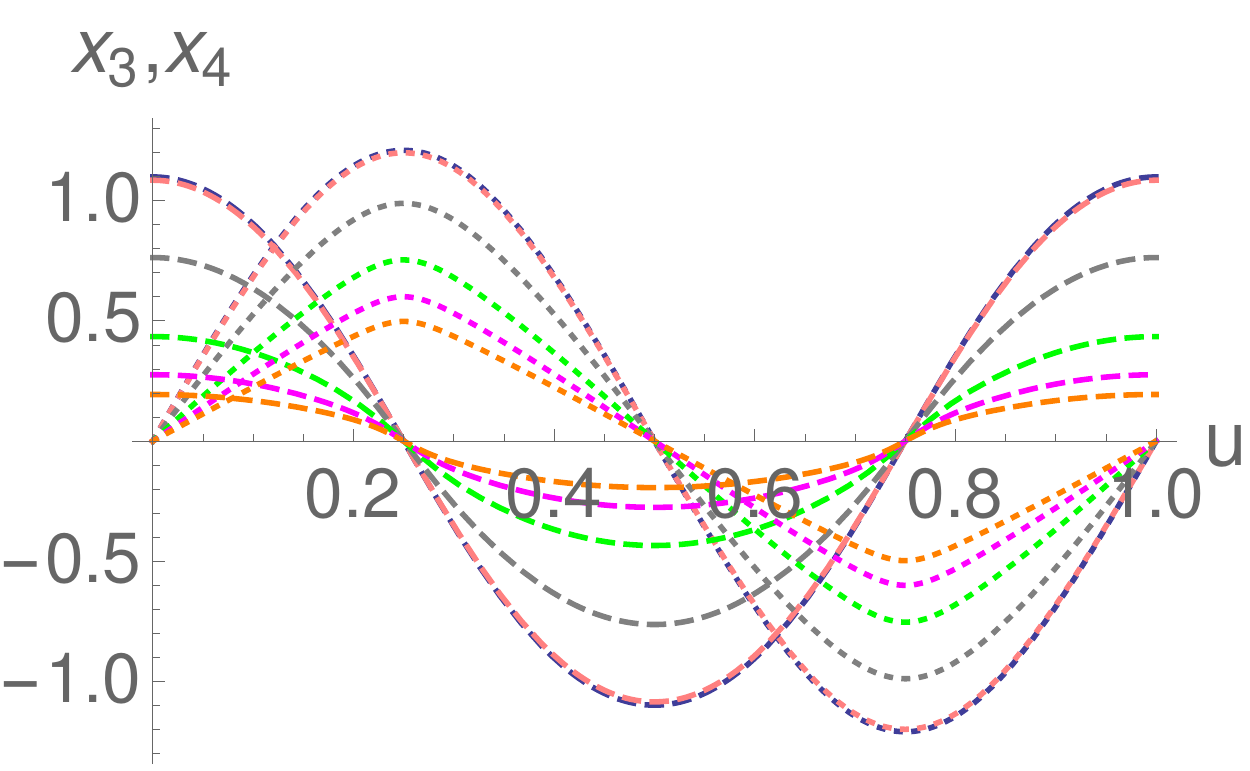}
\caption{Worldline instantons for superimposed temporal sinuosidal field with $\gamma_\omega \in \{0.001,5,10,15,20,25\}$ (from blue, outer path, to orange, inner path). In the right panel the components $x_3$ (dashed) and $x_4$ (dotted) are separately plotted. Remaining field parameters are chosen as $\gamma_k = 0.5$ and $\epsilon \in \{10^{-1},10^{-3},10^{-4}\}$ (from top to bottom).
}
\label{fig:WLIs-sinusoid}
\end{figure}

\begin{figure}[!h]
  \centering
\includegraphics[width=.17\textwidth]{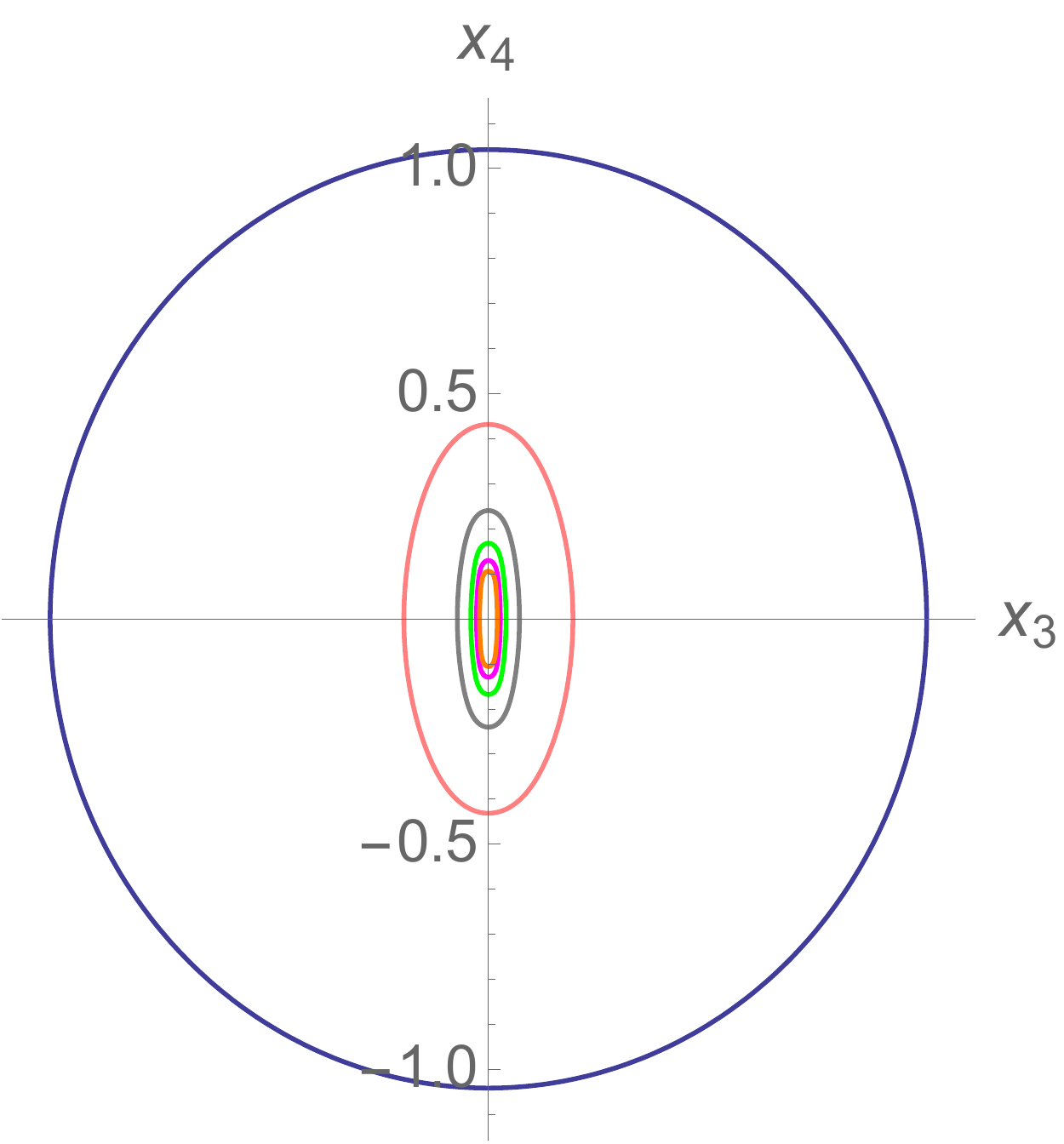}
\includegraphics[width=.30\textwidth]{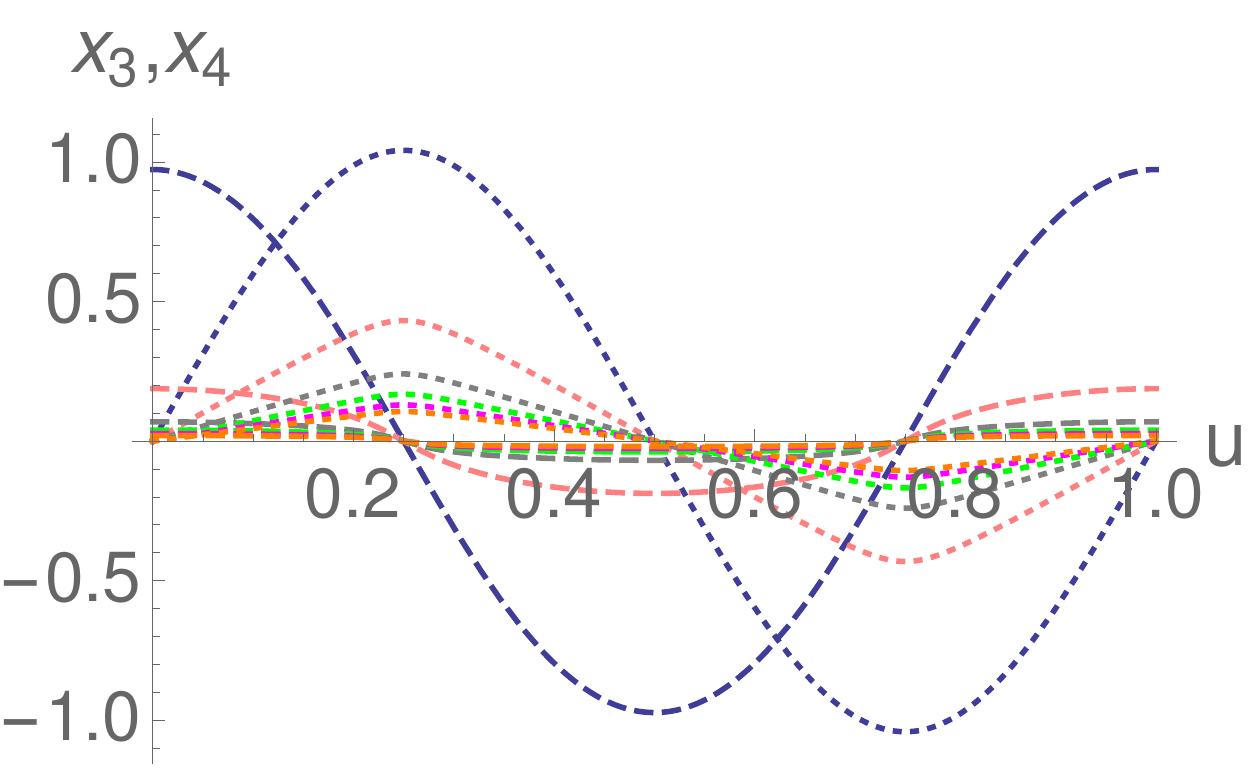}
\includegraphics[width=.17\textwidth]{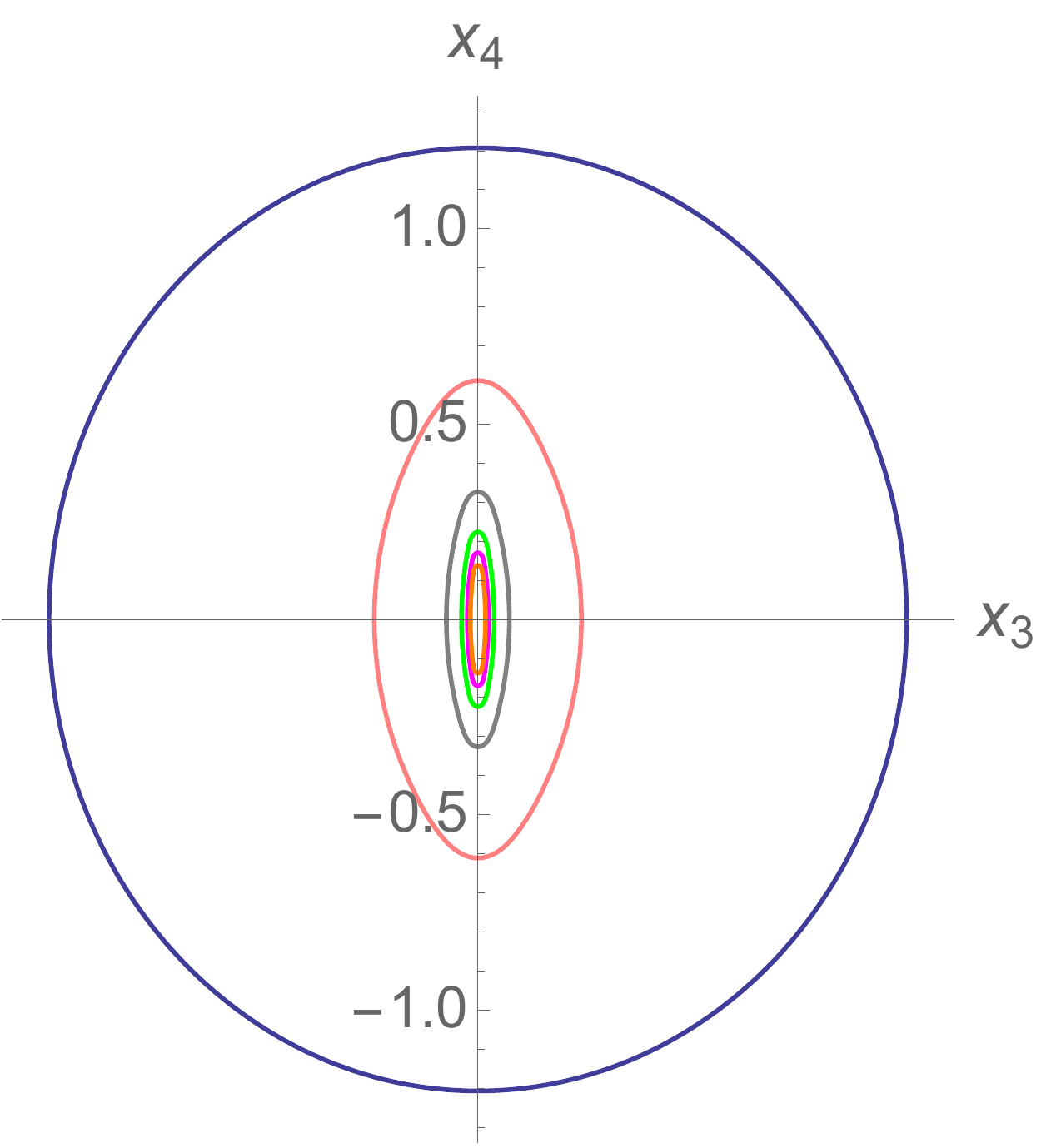}
\includegraphics[width=.30\textwidth]{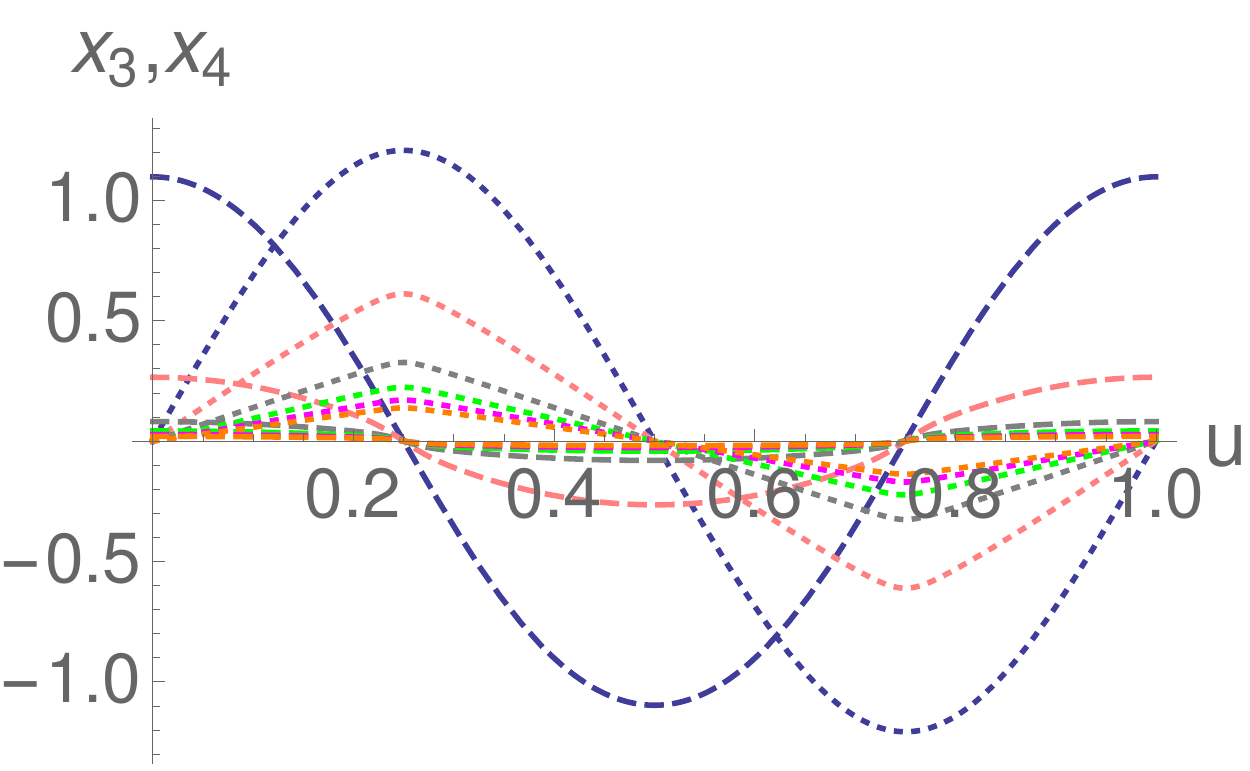}
\includegraphics[width=.17\textwidth]{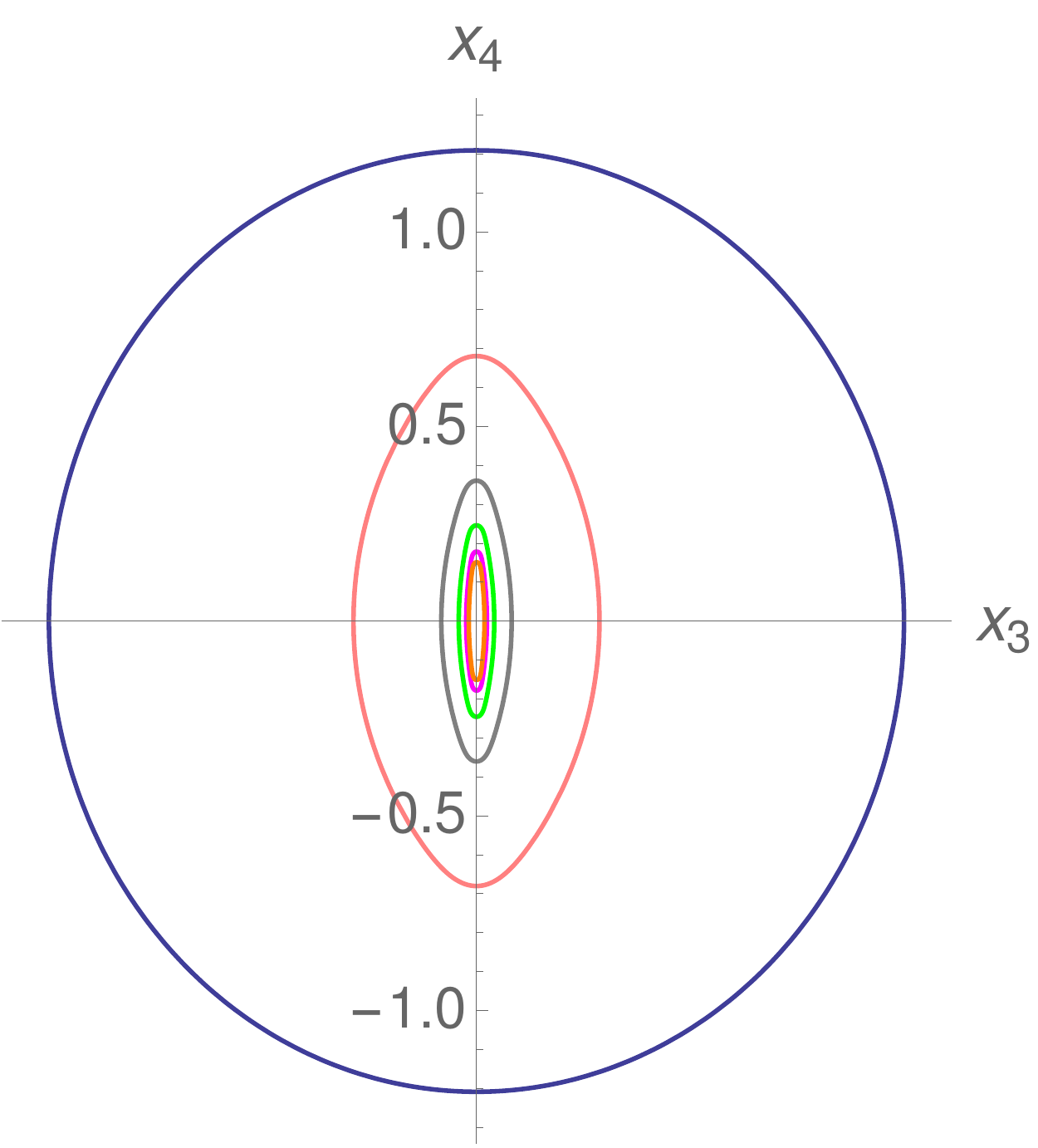}
\includegraphics[width=.30\textwidth]{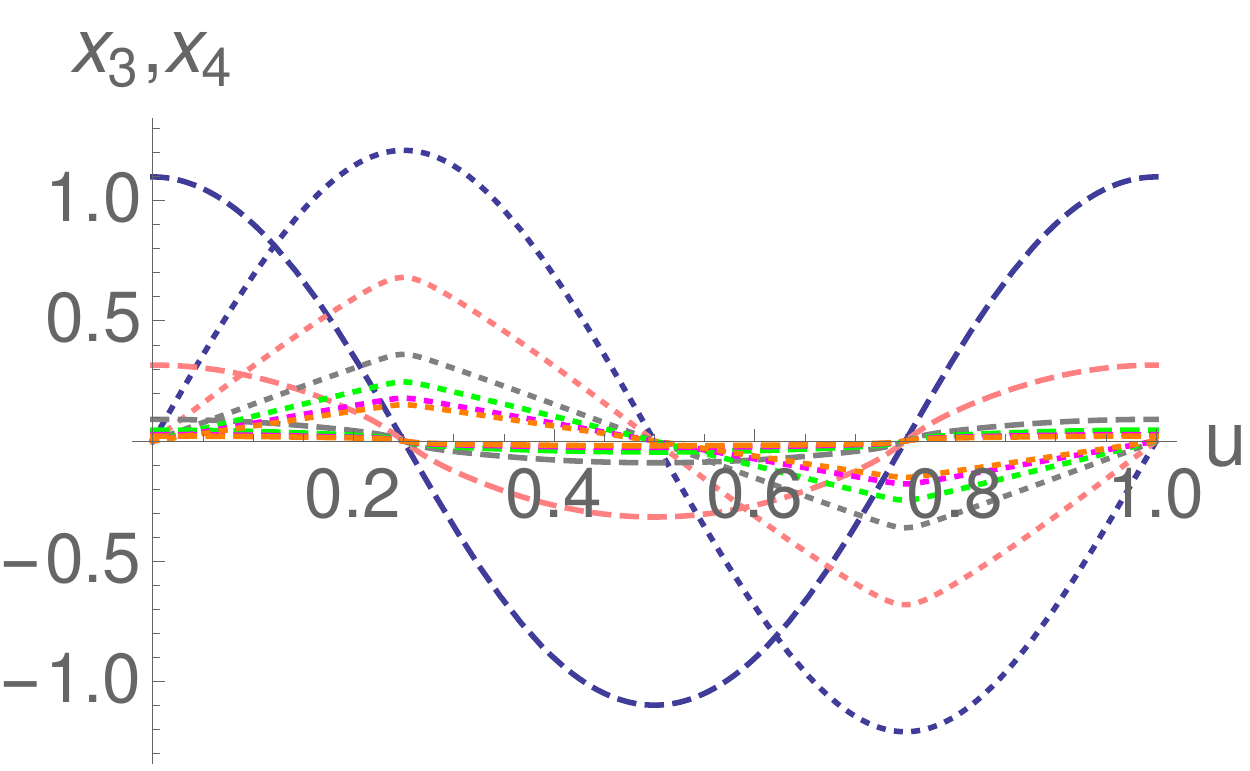}
\caption{Worldline instantons for superimposed temporal Gaussian field with $\gamma_\omega \in \{0.001,5,10,15,20,25\}$ (from blue, outer path, to orange, inner path). In the right panel the components $x_3$ (dashed) and $x_4$ (dotted) are separately plotted. Remaining field parameters are chosen as $\gamma_k = 0.5$ and $\epsilon \in \{10^{-1},10^{-3},10^{-4}\}$ (from top to bottom).
}
\label{fig:WLIs-gaussian}
\end{figure}

Following the strategy described in Sec.~\ref{subsec:restricts}, we find the corresponding instanton paths for any field parameters of interest.
The new results for a spatial Sauter field, superimposed with a time dependent sinusoidal and Gaussian field, are shown in Figs.~\ref{fig:WLIs-sinusoid} and \ref{fig:WLIs-gaussian}, respectively, fixing $\gamma_k = 0.5$ and varying only $\gamma_\omega$ and $\epsilon$ where the
corresponding numerical values are listed in the figure captions.
For backgrounds with a weak temporal Sauter-like dependence,
instantons admit a lens-shape with cusped turning sections \cite{Schneider:2014mla,Akal:2017ilh}, see Fig.~\ref{fig:WLIs-sauter}.
However, for poleless fields, reflections turn out to be softened and the paths curve much smoother.
For $\gamma_\omega \gg 1$ they tend to become increasingly lens-shaped but still remain smoothly curved.
This effect seems to be much stronger for the Gaussian field, cf. Figs.~\ref{fig:WLIs-sinusoid} and \ref{fig:WLIs-gaussian}.
Furthermore, due to the additional $\epsilon$ dependence, there appear significant differences among the paths, independent from the field profile.
The described effects are much more pronounced for the sinusoidal field.

The presence of poles for Sauter-like fields is basically responsible for the $\epsilon$ independence and the seemingly related cusps.
This may explain why the leading order exponential factor in $\mathcal{P}$ can be accurately approached already at $\mathcal{O}(\epsilon)$ in perturbation theory \cite{Torgrimsson:2017pzs}.
Since $\mathcal{W}_0$ does not
feature any $\epsilon$ dependence, at least in the limit $\epsilon \ll 1$, the same exponent has to apply at any higher order in $\epsilon$. Therefore, the first order contribution stemming from the weak field should indeed be capable to approximate $\mathcal{W}_0$. Note that we treat the background nonperturbatively.

For time dependent fields, such as of sinusoidal and Gaussian type, poles are not present.
In these cases, since $\mathcal{W}_0$ does in general depend on $\epsilon$, we may expect different exponents in the perturbative expansion of $\mathcal{P}$.
Hence, the effective reflection picture already elucidates the relevance of higher orders in $\epsilon$ for poleless fields as highlighted in \cite{Torgrimsson:2017pzs}.
Interestingly, super Gaussian fields of the form
\begin{align}
E_\omega e^{-(\omega t)^{4N+2}},\ N \in \mathbb{N}
\label{eq:super-gauss}
\end{align}
may reveal some striking properties. 
Namely, in a purely temporal setup, the $\epsilon$ dependence of $\mathcal{W}_0$ becomes increasingly suppressed for weak super Gaussians of the form \eqref{eq:super-gauss}. For $N \rightarrow \infty$ the action $\mathcal{W}_0$ is even expected to converge to the Lorentzian limit, see e.g. \cite{Akal:2017ilh}, although such setups crucially differ in Minkowski spacetime.
These aspects will be studied in \cite{Akal-sg:2017}.

Coming back to the worldline instantons in Figs.~\ref{fig:WLIs-sinusoid} and \ref{fig:WLIs-gaussian},
the advantage for treating the system with the help of $\pmb{\Omega}_2$ and rearranging the starting points is clearly reflected.
The root finding works very robustly and provides the correct paths independently from the chosen field parameters
which was not possible with $\pmb{\Omega}_1$. Additional constraints resulting from the underlying instanton symmetry improve the root finding routine substantially.
\subsection{Stationary actions}
\label{subsec:W0}
\begin{figure}[!h]
  \centering
\includegraphics[width=.4\textwidth]{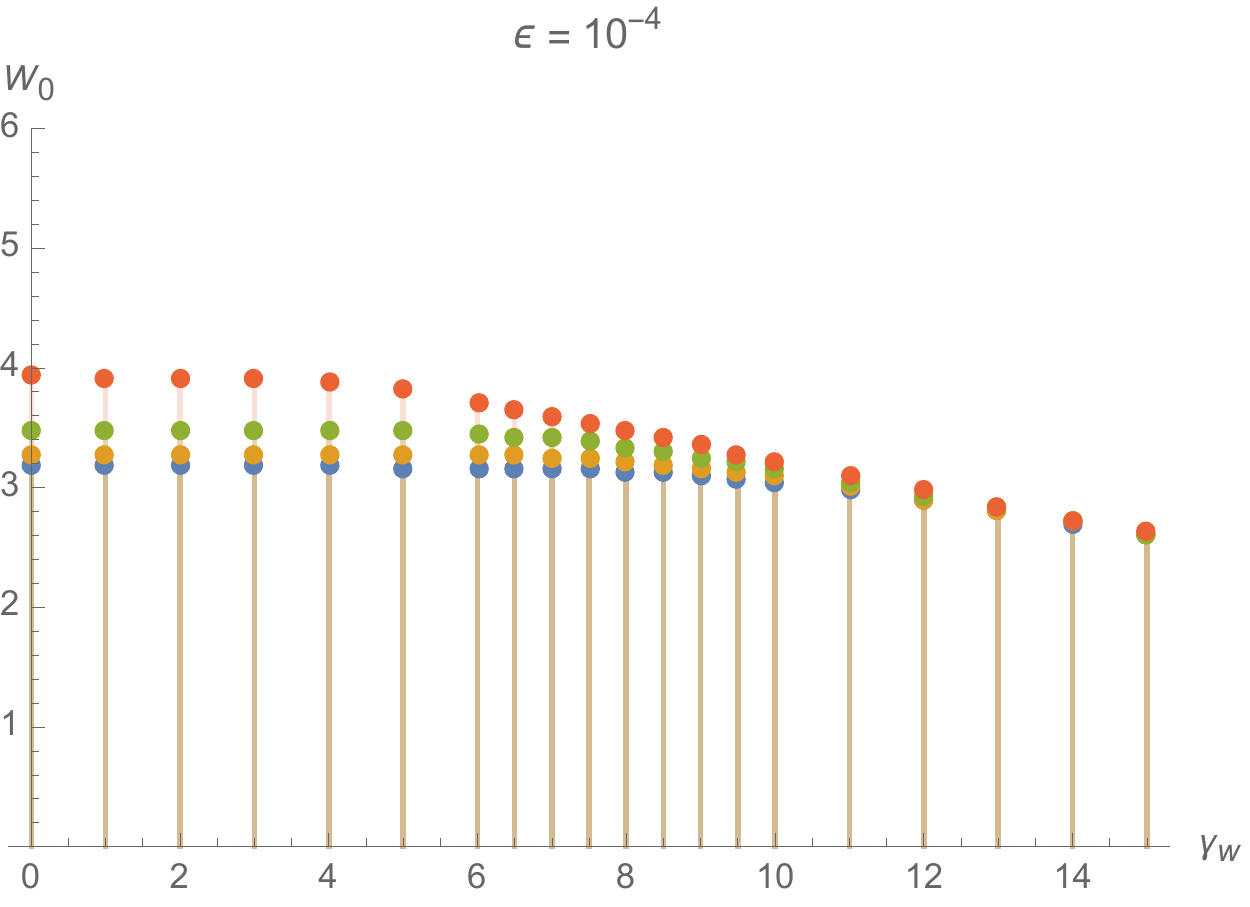}\\
\vspace{0.5cm}
\includegraphics[width=.49\textwidth]{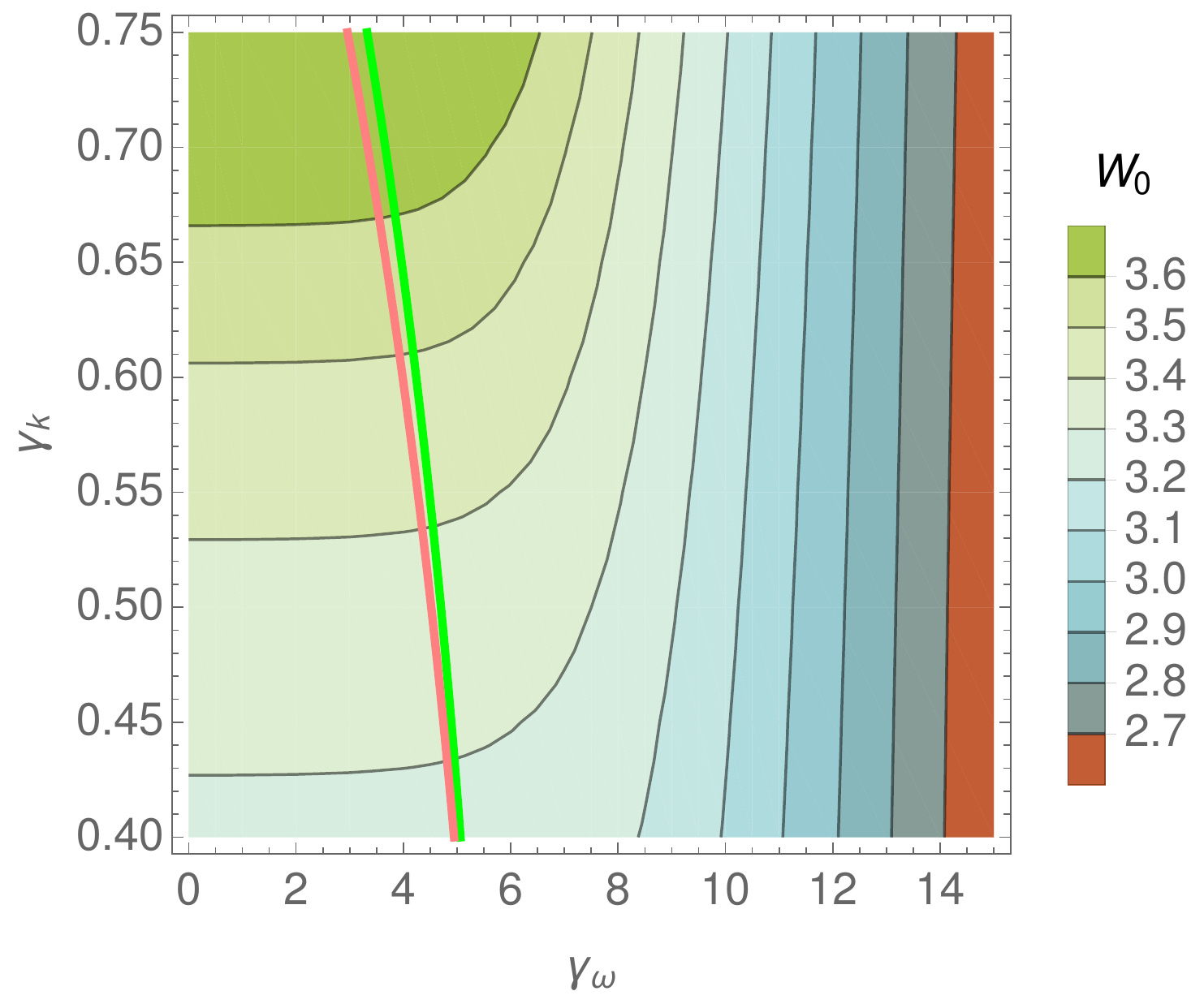}
\caption{Top panel: $\mathcal{W}_0\ [E_\mathrm{S}/ E_k]$ plotted versus $\gamma_\omega$ for the temporal sinusoidal field for fixed $\gamma_k \in \{0.2,0.4,0.6,0.8\}$ (from blue to red) and $\epsilon = 10^{-4}$. Bottom panel: $\mathcal{W}_0$ is depicted as a contour plot. The thick lines are the analytically predicted $\gamma_\omega^\mathrm{crit}$ from \eqref{eq:crit-gamma_omega} with (pink) and without (green) the modified field strength parameter $\tilde\epsilon$, see  \eqref{eq:epsilon-corr}, included.}
\label{fig:W0-Sinusoid}
\end{figure}

\begin{figure}[!h]
  \centering
\includegraphics[width=.4\textwidth]{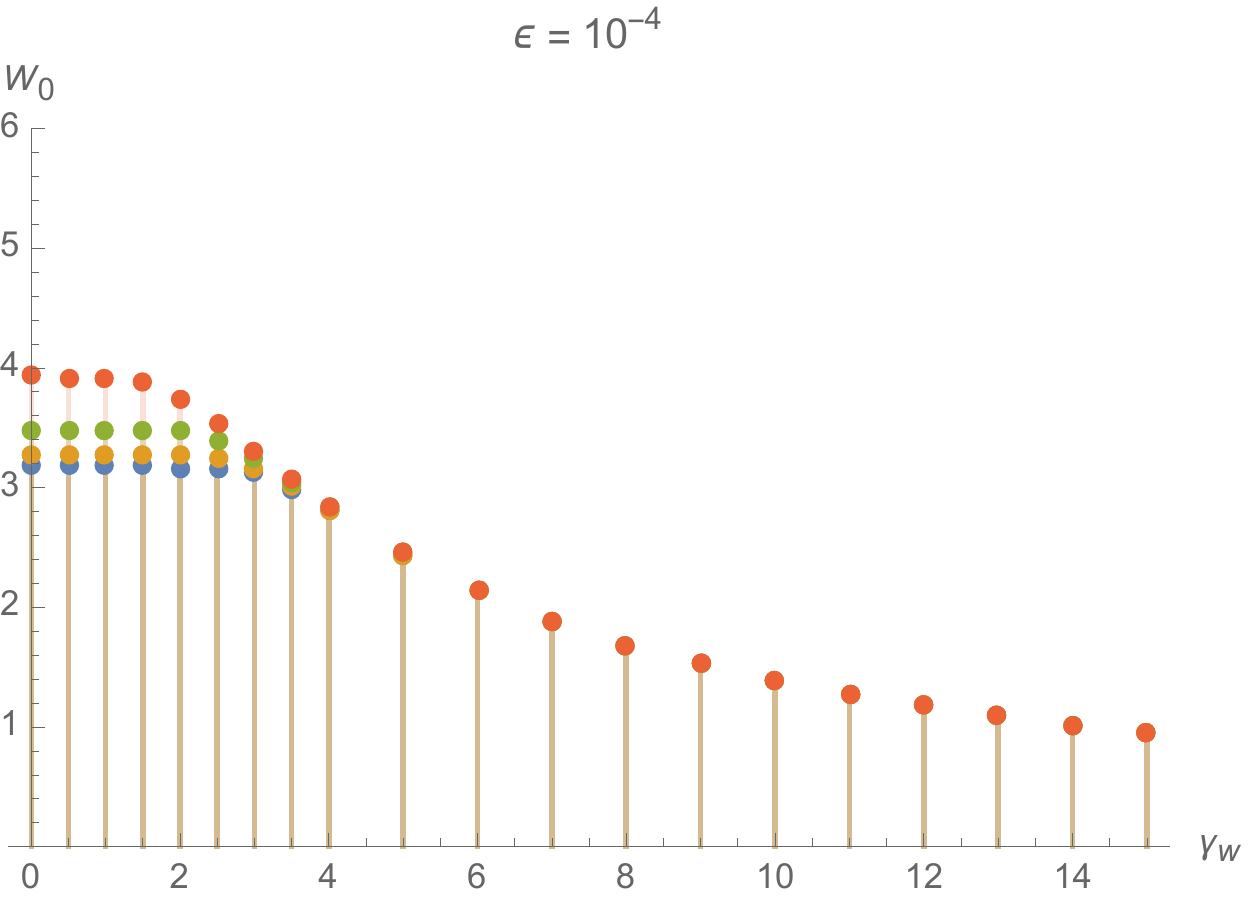}\\
\vspace{0.5cm}
\includegraphics[width=.49\textwidth]{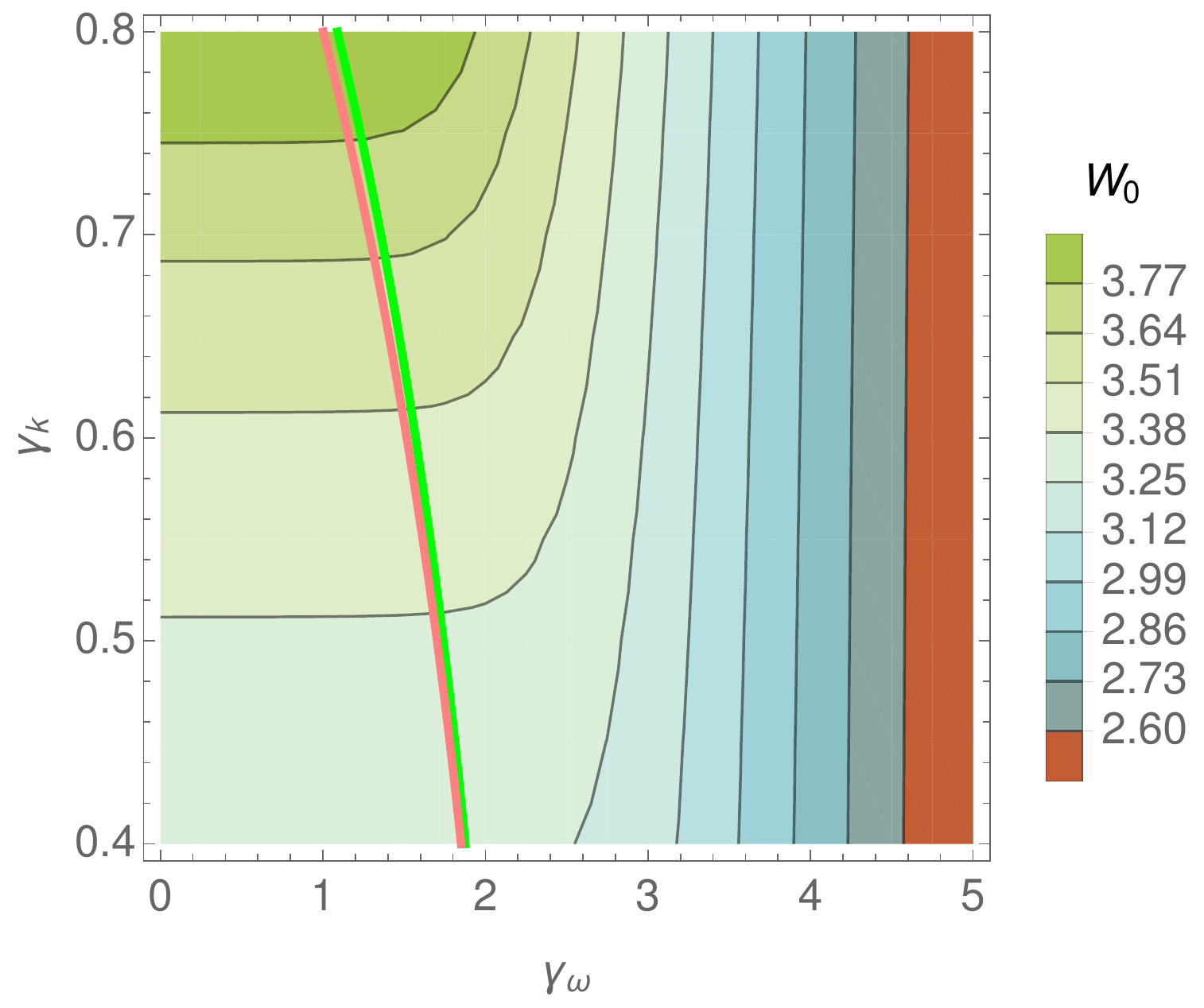}
\caption{Top panel: $\mathcal{W}_0\ [E_\mathrm{S}/ E_k]$ plotted versus $\gamma_\omega$ for the temporal Gaussian field for fixed $\gamma_k \in \{0.2,0.4,0.6,0.8\}$ (from blue to red) and $\epsilon = 10^{-4}$. Bottom panel: $\mathcal{W}_0$ is depicted as a contour plot. The thick lines are the analytically predicted $\gamma_\omega^\mathrm{crit}$ from \eqref{eq:crit-gamma_omega} with (pink) and without (green) the modified field strength parameter $\tilde\epsilon$, see  \eqref{eq:epsilon-corr}, included.}
\label{fig:W0-Gaussian}
\end{figure}

What still remains is the computation of the stationary worldline action, $\mathcal{W}_0$. This can be directly performed
following the previous recipe in Sec.~\ref{subsec:restricts}.
First, we will numerically find the worldline instantons for a set of parameters
and take the data afterwards to evaluate $\mathcal{W}$ on these paths.
For this, let us fix the field strength ratio as $\epsilon = 10^{-4}$. The spatial as well as Keldysh parameter is varied in a convenient range.
The results for $\mathcal{W}_0$ are depicted in Figs.~\ref{fig:W0-Sinusoid} and \ref{fig:W0-Gaussian}, respectively.
In the top panels, $\mathcal{W}_0$ has been plotted versus $\gamma_\omega$ for different $\gamma_k$ listed in the corresponding captions. In both cases one finds that $\mathcal{W}_0 \uparrow$ if $\gamma_k \uparrow$ as long as
$\gamma_\omega \ll \gamma_\omega^\mathrm{crit}$.
If the weak field starts to assist, i.e. $\gamma_\omega > \gamma_\omega^\mathrm{crit}$,
we find $\mathcal{W}_0 \downarrow$ for $\gamma_\omega \uparrow$. For $\gamma_\omega \gg \gamma_\omega^\mathrm{crit}$ the different curves converge to a single curve which one would obtain for $\gamma_k = 0$. This is in agreement with our expectation, since a static spatial field would provide the largest contribution to the effective total field strength. Interestingly, the critical threshold for the weak sinusoidal field applies much later compared to the Gaussian case. Moreover, for the former field all curves drop much slower for temporal Keldysh parameters $\gamma_\omega > \gamma_\omega^\mathrm{crit}$. This is consistent with recent observations in \cite{Akal:2017ilh}. There, such differences have been argued to be caused by the relatively large effective reflection point. Indeed, this has been presumed to be the key reason why a weak time dependent sinusoidal field assists less than a Sauter pulse for which the reflection point is much smaller and, even more important, $\epsilon$ independent \cite{Schutzhold:2008pz,Schneider:2016vrl,Schneider:2014mla,Linder:2015vta,Akal:2017ilh}. As a consequence, in the latter case, worldline instantons are reflected and squeezed already for relatively small $\gamma_\omega$
leading to the mentioned faster decrease of $\mathcal{W}_0$.
The effective reflection picture, as discussed in \cite{Akal:2017ilh}, helps to understand such differences in a quite intuitive way in terms of instanton reflections.

In the bottom panels of Figs.~\ref{fig:W0-Sinusoid} and \ref{fig:W0-Gaussian}, the separate curves are combined in a contour plot where $\mathcal{W}_0$ is plotted versus $\gamma_k$ and $\gamma_\omega$. The color maps on the RHS are scaled according to the numerical values of $\mathcal{W}_0$.
The previously described trends are again clearly reflected.
However, in addition, we have now included the analytically approximated critical threshold $\gamma_\omega^\mathrm{crit}$ from \eqref{eq:crit-gamma_omega} as well. The difference between the shown two critical curves, one in green and the other in pink, is that the former has been generated without incorporating the modified field strength parameter $\tilde \epsilon$ from \eqref{eq:epsilon-corr} in \eqref{eq:crit-gamma_omega}. For $\gamma_k \uparrow$ the
curve with $\tilde\epsilon$ included, is much more accurate being in agreement with the discussion in Sec.~\ref{subsec:ana-approx}.
To the right of this critical curve, we find strong
evidence for
dynamical assistance
indicated by the strongly bent gray, solid contour lines.
Hence, the analytical approximations match very well with the exact numerical results.
Such remarkable agreements suggest that the present approach serves as
an efficient way to get some analytical insights even in cases with such complex backgrounds.
\section{Strong time dependence}
\label{sec:strong-T}
\begin{figure}[h]
  \centering
\includegraphics[width=.18\textwidth]{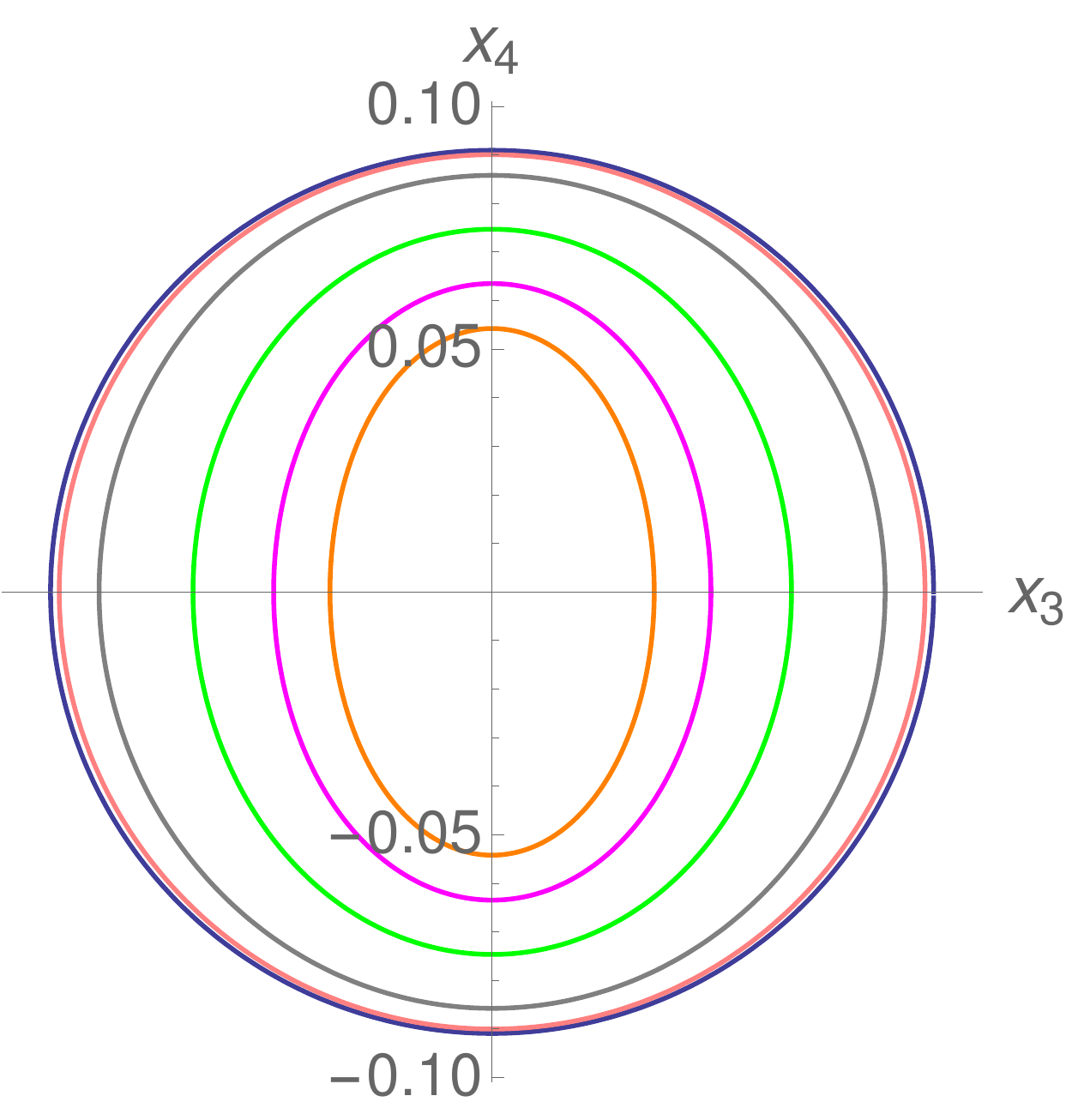}
\includegraphics[width=.28\textwidth]{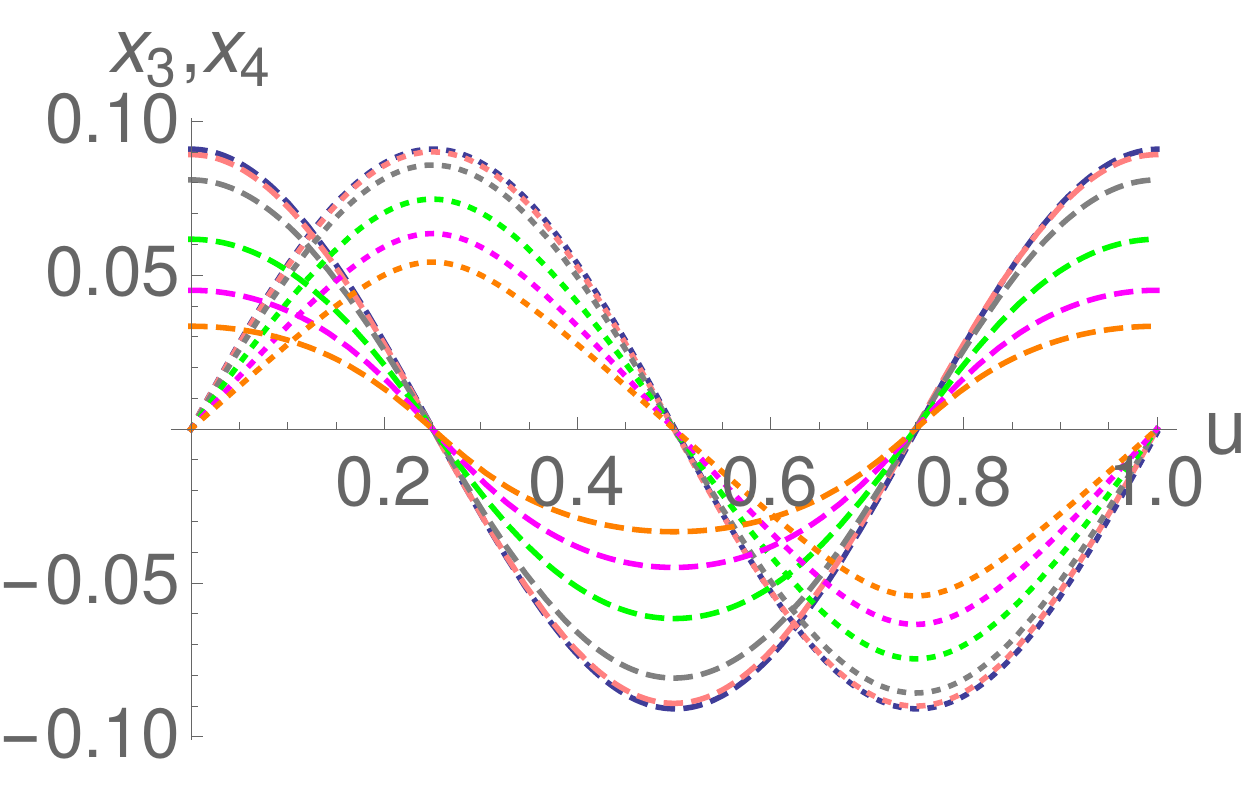}
\caption{Worldline instantons for a spatial Sauter field combined with a stronger temporal Sauter field. The field parameters are $\gamma_k = 0.5$, $\epsilon = 10$ and $\gamma_\omega \in \{0.001,2,5,10,15,20\}$ (blue, outer path, to orange, inner path). The two components $x_3$ (dashed) and $x_4$ (dotted) are separately plotted in the right panel. The values are given in units of $[m/ E_k]$.}
\label{fig:WLIs-sauter-largeEps}
\end{figure}

In an electric background being too localised in space, $\gamma_k \geq 1$, tunnelling of virtual dipole pairs is not possible for $\epsilon \ll 1$ and $\gamma_\omega \rightarrow 0$. This corresponds to the non-existence of a periodic path in spacetime, see e.g. \cite{Dunne:2005sx,Gies:2005bz}.
However, for the present type of backgrounds this effect will be absent if $\epsilon > 1$.
We assume the background to be the linear combination of two Sauter pulses. 
The numerical computation strategy is the same as discussed in Sec.~\ref{subsec:restricts} which works very robust even for the present purpose.
The obtained worldline instantons are depicted in the left panel of Fig.~\ref{fig:WLIs-sauter-largeEps}, whereas in the right panel, both space and time components, $x_3$ and $x_4$, are plotted separately.
Chosen field parameters are given in the figure caption.
In case of $\gamma_\omega \uparrow$ the instanton paths tend to shrink smoothly, means no appearance of discontinuities in form of cusped turning sections.
More importantly, for $\gamma_k \uparrow$ instantons are real \cite{Dunne:2005sx}, since the dominant contribution comes from the stronger temporal dependence, see discussion in Sec.~\ref{subsec:ana-approx}.
Thus, there will be no additional instanton reflections and, consequently, no dynamical assistance.
For $\gamma_k \gg 1$ the spatial contribution will become increasingly negligible.
As soon as $\gamma_k \rightarrow 0$, the strength of the spatial Sauter field approaches its peak value and will maximally contribute to the total effective field strength. This will shrink the instanton paths even more.
\section{Conclusion}
\label{sec:conc}
We have investigated the tunnelling process of virtual pairs from the quantum vacuum in the presence of certain 
multidimensional ($1+1$) electric backgrounds which depend on space as well as on time.
Going beyond the case of two linearly combined Sauter pulses \cite{Schneider:2014mla}
we have considered a weak time dependence of sinusoidal and Gaussian type, respectively, which do not have poles in the complex plane.
Using the worldline formalism,
the resulting background has been treated nonperturbatively.
The underlying equations have been simplified by applying certain effective critical points, recently presented in \cite{Akal:2017ilh}.
On this basis we could analytically predict a threshold $\gamma_\omega^\mathrm{crit}$ for the temporal inhomogeneity $\gamma_\omega$ depending on both the 
field strength ratio $\epsilon$ and the spatial inhomogeneity $\gamma_k$.

We have set appropriate initial conditions and applied additional symmetry constraints
present due to the assumed background structure. 
These steps allowed us to find the corresponding worldline instantons for any parameters of interest.
Using these closed paths in spacetime, the leading order exponential factors for both backgrounds have been computed, finding a large dynamical enhancement in general.
We have seen that below the predicted threshold $\gamma_\omega^\mathrm{crit}$ there is no substantial contribution from the weak term.

Furthermore, we have found that such backgrounds lead in general 
to a smaller enhancement compared to the case with a Sauter-like time variation.
That is due to the fact that for $\gamma_k \rightarrow 0$, the limit where the spatial term maximally contributes to the delocalisation of the virtual pair,
the critical threshold $\gamma_\omega^\mathrm{crit}$ becomes relatively large.
This effect is much more likely in the oscillatory sinusoidal case.
On the other hand, for $\gamma_k \rightarrow 1$ the width of the spatial Sauter pulse decreases towards the critical Compton region 
with the consequence that $\gamma_\omega^\mathrm{crit} \rightarrow 0$. In this case even arbitrarily small time variations are not negligible.
The tunnelling rate, however, decreases due to a large exponential suppression
which slows down the enhancement even more compared to the Sauter case. 

Finally, we also discussed the $1+1$ dimensional double Sauter background for $\epsilon > 1$ and studied the main differences we have found for the instanton paths compared to earlier studies with $\epsilon \ll 1$.
 
\section{Acknowledgments}
We acknowledge the support of the Colloborative Research Center SFB 676 \textit{Particles, Strings and the Early Universe} of the DFG.

\appendix
\section{Critical Keldysh parameter}
\label{app:eff-strength}
The critical temporal Keldysh parameter $\gamma_\omega^\mathrm{crit}$ in case of $\gamma_k > 0$ requires a modification of the (effective) spatial field strength.
This modification goes back to the observation in \cite{Akal:2017ilh} where
\begin{align}
  \gamma_k \uparrow\  \Rightarrow \Delta \uparrow\ \Rightarrow \gamma_\omega^\mathrm{crit} \downarrow
\end{align}
holds in general, cf. \eqref{eq:x4crit}.
The corresponding value can be obtained by taking $\mathrm{max}\{x_3\}$ which minimizes the field strength for a fixed $\gamma_k$. This maximum may be computed only for the spatial Sauter field, since we are interested in the critical threshold where the contribution of the temporal field may still be assumed as negligible. Note that the enormous enhancement applies for values above the threshold $\gamma_\omega^\mathrm{crit}$.
In this case the additional contribution will for sure decrease $\mathrm{max}\{x_3\}$ which is, however, irrelevant for the present purpose.
For the spatial Sauter field the exact instanton solution reads
\eqnsplit{
x_3(u) &= \frac{1}{\gamma_k} \mathrm{arcsinh}\left( \frac{\gamma_k}{\sqrt{1-\gamma_k^2}} \cos(2 \pi n u) \right),\\
x_4(u) &= \frac{1}{\gamma_k\sqrt{1 - \gamma_k^2}} \mathrm{arcsin}\left( \gamma_k \sin(2 \pi n u) \right).
}
Taking the leading worldline instanton with winding number $n=1$ \cite{Affleck:1981bma,Dunne:2005sx}, the maximum of the spatial component is reached at (rescaled) proper-time $u=0$,
\eqnsplit{
 x_{3,\text{max}} = \frac{1}{\gamma_k} \mathrm{arcsinh}\left(\frac{\gamma_k}{\sqrt{1-\gamma_k^2}}\right),
}
which subsequently results in
\eqnsplit{
  \underset{\text{fixed}\ k,E_k}{\text{min}} &\left\{ E_k\mathrm{sech}^2(k x_3) \right\}
  = E_k\mathrm{sech}^2(k x_{3,\text{max}})\\
  &= E_k \mathrm{sech}^2\left(\mathrm{arcsinh}\left(\frac{\gamma_k}{\sqrt{1-\gamma_k^2}}\right)\right).
}
Hence, the effective field strength ratio for the general case $0 \leq \gamma_k < 1$ takes the form
\begin{align}
 \tilde \epsilon = \epsilon \mathrm{cosh}^2\left(\mathrm{arcsinh}\left(\frac{\gamma_k}{\sqrt{1-\gamma_k^2}}\right)\right)
\end{align}
which, accordingly, has to be plugged into \eqref{eq:crit-gamma_omega}, replacing the initial parameter $\epsilon$.

\bibliography{article_bib}

\begin{thebibliography}{74}%
\makeatletter
\providecommand \@ifxundefined [1]{%
 \@ifx{#1\undefined}
}%
\providecommand \@ifnum [1]{%
 \ifnum #1\expandafter \@firstoftwo
 \else \expandafter \@secondoftwo
 \fi
}%
\providecommand \@ifx [1]{%
 \ifx #1\expandafter \@firstoftwo
 \else \expandafter \@secondoftwo
 \fi
}%
\providecommand \natexlab [1]{#1}%
\providecommand \enquote  [1]{``#1''}%
\providecommand \bibnamefont  [1]{#1}%
\providecommand \bibfnamefont [1]{#1}%
\providecommand \citenamefont [1]{#1}%
\providecommand \href@noop [0]{\@secondoftwo}%
\providecommand \href [0]{\begingroup \@sanitize@url \@href}%
\providecommand \@href[1]{\@@startlink{#1}\@@href}%
\providecommand \@@href[1]{\endgroup#1\@@endlink}%
\providecommand \@sanitize@url [0]{\catcode `\\12\catcode `\$12\catcode
  `\&12\catcode `\#12\catcode `\^12\catcode `\_12\catcode `\%12\relax}%
\providecommand \@@startlink[1]{}%
\providecommand \@@endlink[0]{}%
\providecommand \url  [0]{\begingroup\@sanitize@url \@url }%
\providecommand \@url [1]{\endgroup\@href {#1}{\urlprefix }}%
\providecommand \urlprefix  [0]{URL }%
\providecommand \Eprint [0]{\href }%
\providecommand \doibase [0]{http://dx.doi.org/}%
\providecommand \selectlanguage [0]{\@gobble}%
\providecommand \bibinfo  [0]{\@secondoftwo}%
\providecommand \bibfield  [0]{\@secondoftwo}%
\providecommand \translation [1]{[#1]}%
\providecommand \BibitemOpen [0]{}%
\providecommand \bibitemStop [0]{}%
\providecommand \bibitemNoStop [0]{.\EOS\space}%
\providecommand \EOS [0]{\spacefactor3000\relax}%
\providecommand \BibitemShut  [1]{\csname bibitem#1\endcsname}%
\let\auto@bib@innerbib\@empty
\bibitem [{\citenamefont {Heisenberg}\ and\ \citenamefont
  {Euler}(1936)}]{Heisenberg:1935qt}%
  \BibitemOpen
  \bibfield  {author} {\bibinfo {author} {\bibfnamefont {W.}~\bibnamefont
  {Heisenberg}}\ and\ \bibinfo {author} {\bibfnamefont {H.}~\bibnamefont
  {Euler}},\ }\href {\doibase 10.1007/BF01343663} {\bibfield  {journal}
  {\bibinfo  {journal} {Z. Phys.}\ }\textbf {\bibinfo {volume} {98}},\ \bibinfo
  {pages} {714} (\bibinfo {year} {1936})},\ \Eprint
  {http://arxiv.org/abs/physics/0605038} {arXiv:physics/0605038 [physics]}
  \BibitemShut {NoStop}%
\bibitem [{\citenamefont {Schwinger}(1951)}]{Schwinger:1951nm}%
  \BibitemOpen
  \bibfield  {author} {\bibinfo {author} {\bibfnamefont {J.~S.}\ \bibnamefont
  {Schwinger}},\ }\href {\doibase 10.1103/PhysRev.82.664} {\bibfield  {journal}
  {\bibinfo  {journal} {Phys. Rev.}\ }\textbf {\bibinfo {volume} {82}},\
  \bibinfo {pages} {664} (\bibinfo {year} {1951})}\BibitemShut {NoStop}%
\bibitem [{\citenamefont {Dittrich}\ and\ \citenamefont
  {Gies}(2000)}]{Dittrich:2000zu}%
  \BibitemOpen
  \bibfield  {author} {\bibinfo {author} {\bibfnamefont {W.}~\bibnamefont
  {Dittrich}}\ and\ \bibinfo {author} {\bibfnamefont {H.}~\bibnamefont
  {Gies}},\ }\href {\doibase 10.1007/3-540-45585-X} {\bibfield  {journal}
  {\bibinfo  {journal} {Springer Tracts Mod. Phys.}\ }\textbf {\bibinfo
  {volume} {166}},\ \bibinfo {pages} {1} (\bibinfo {year} {2000})}\BibitemShut
  {NoStop}%
\bibitem [{\citenamefont {Allor}\ \emph {et~al.}(2008)\citenamefont {Allor},
  \citenamefont {Cohen},\ and\ \citenamefont {McGady}}]{Allor:2007ei}%
  \BibitemOpen
  \bibfield  {author} {\bibinfo {author} {\bibfnamefont {D.}~\bibnamefont
  {Allor}}, \bibinfo {author} {\bibfnamefont {T.~D.}\ \bibnamefont {Cohen}}, \
  and\ \bibinfo {author} {\bibfnamefont {D.~A.}\ \bibnamefont {McGady}},\
  }\href {\doibase 10.1103/PhysRevD.78.096009} {\bibfield  {journal} {\bibinfo
  {journal} {Phys. Rev.}\ }\textbf {\bibinfo {volume} {D78}},\ \bibinfo {pages}
  {096009} (\bibinfo {year} {2008})},\ \Eprint {http://arxiv.org/abs/0708.1471}
  {arXiv:0708.1471 [cond-mat.mes-hall]} \BibitemShut {NoStop}%
\bibitem [{\citenamefont {Katsnelson}\ and\ \citenamefont
  {Volovik}(2012)}]{Katsnelson:2012cz}%
  \BibitemOpen
  \bibfield  {author} {\bibinfo {author} {\bibfnamefont {M.~I.}\ \bibnamefont
  {Katsnelson}}\ and\ \bibinfo {author} {\bibfnamefont {G.~E.}\ \bibnamefont
  {Volovik}},\ }\href {\doibase 10.1134/S0021364012080061} {\bibfield
  {journal} {\bibinfo  {journal} {JETP Lett.}\ }\textbf {\bibinfo {volume}
  {95}},\ \bibinfo {pages} {411} (\bibinfo {year} {2012})},\ \bibinfo {note}
  {[Pisma Zh. Eksp. Teor. Fiz.95,457(2012)]},\ \Eprint
  {http://arxiv.org/abs/1203.1578} {arXiv:1203.1578 [cond-mat.str-el]}
  \BibitemShut {NoStop}%
\bibitem [{\citenamefont {Zubkov}(2012)}]{Zubkov:2012ht}%
  \BibitemOpen
  \bibfield  {author} {\bibinfo {author} {\bibfnamefont {M.~A.}\ \bibnamefont
  {Zubkov}},\ }\href {\doibase 10.1134/S0021364012090135} {\bibfield  {journal}
  {\bibinfo  {journal} {Pisma Zh. Eksp. Teor. Fiz.}\ }\textbf {\bibinfo
  {volume} {95}},\ \bibinfo {pages} {540} (\bibinfo {year} {2012})},\ \Eprint
  {http://arxiv.org/abs/1204.0138} {arXiv:1204.0138 [hep-ph]} \BibitemShut
  {NoStop}%
\bibitem [{\citenamefont {Fillion-Gourdeau}\ and\ \citenamefont
  {MacLean}(2015)}]{Fillion-Gourdeau:2015dga}%
  \BibitemOpen
  \bibfield  {author} {\bibinfo {author} {\bibfnamefont {F.}~\bibnamefont
  {Fillion-Gourdeau}}\ and\ \bibinfo {author} {\bibfnamefont {S.}~\bibnamefont
  {MacLean}},\ }\href {\doibase 10.1103/PhysRevB.92.035401} {\bibfield
  {journal} {\bibinfo  {journal} {Phys. Rev.}\ }\textbf {\bibinfo {volume}
  {B92}},\ \bibinfo {pages} {035401} (\bibinfo {year} {2015})}\BibitemShut
  {NoStop}%
\bibitem [{\citenamefont {Linder}\ and\ \citenamefont
  {Sch{\"u}tzhold}(2015)}]{Linder:2015fba}%
  \BibitemOpen
  \bibfield  {author} {\bibinfo {author} {\bibfnamefont {M.~F.}\ \bibnamefont
  {Linder}}\ and\ \bibinfo {author} {\bibfnamefont {R.}~\bibnamefont
  {Sch{\"u}tzhold}},\ }\href@noop {} {\  (\bibinfo {year} {2015})},\ \Eprint
  {http://arxiv.org/abs/1503.07108} {arXiv:1503.07108 [cond-mat.mes-hall]}
  \BibitemShut {NoStop}%
\bibitem [{\citenamefont {Akal}\ \emph {et~al.}(2016)\citenamefont {Akal},
  \citenamefont {Egger}, \citenamefont {M{\"u}ller},\ and\ \citenamefont
  {Villalba-Ch{\'a}vez}}]{Akal:2016stu}%
  \BibitemOpen
  \bibfield  {author} {\bibinfo {author} {\bibfnamefont {I.}~\bibnamefont
  {Akal}}, \bibinfo {author} {\bibfnamefont {R.}~\bibnamefont {Egger}},
  \bibinfo {author} {\bibfnamefont {C.}~\bibnamefont {M{\"u}ller}}, \ and\
  \bibinfo {author} {\bibfnamefont {S.}~\bibnamefont {Villalba-Ch{\'a}vez}},\
  }\href {\doibase 10.1103/PhysRevD.93.116006} {\bibfield  {journal} {\bibinfo
  {journal} {Phys. Rev.}\ }\textbf {\bibinfo {volume} {D93}},\ \bibinfo {pages}
  {116006} (\bibinfo {year} {2016})},\ \Eprint
  {http://arxiv.org/abs/1602.08310} {arXiv:1602.08310 [hep-ph]} \BibitemShut
  {NoStop}%
\bibitem [{\citenamefont {Fillion-Gourdeau}\ \emph {et~al.}(2016)\citenamefont
  {Fillion-Gourdeau}, \citenamefont {Gagnon}, \citenamefont {Lefebvre},\ and\
  \citenamefont {MacLean}}]{Fillion-Gourdeau:2016izx}%
  \BibitemOpen
  \bibfield  {author} {\bibinfo {author} {\bibfnamefont {F.}~\bibnamefont
  {Fillion-Gourdeau}}, \bibinfo {author} {\bibfnamefont {D.}~\bibnamefont
  {Gagnon}}, \bibinfo {author} {\bibfnamefont {C.}~\bibnamefont {Lefebvre}}, \
  and\ \bibinfo {author} {\bibfnamefont {S.}~\bibnamefont {MacLean}},\ }\href
  {\doibase 10.1103/PhysRevB.94.125423} {\bibfield  {journal} {\bibinfo
  {journal} {Phys. Rev.}\ }\textbf {\bibinfo {volume} {B94}},\ \bibinfo {pages}
  {125423} (\bibinfo {year} {2016})}\BibitemShut {NoStop}%
\bibitem [{\citenamefont {Abramchuk}\ and\ \citenamefont
  {Zubkov}(2016)}]{Abramchuk:2016afc}%
  \BibitemOpen
  \bibfield  {author} {\bibinfo {author} {\bibfnamefont {R.~A.}\ \bibnamefont
  {Abramchuk}}\ and\ \bibinfo {author} {\bibfnamefont {M.~A.}\ \bibnamefont
  {Zubkov}},\ }\href {\doibase 10.1103/PhysRevD.94.116012} {\bibfield
  {journal} {\bibinfo  {journal} {Phys. Rev.}\ }\textbf {\bibinfo {volume}
  {D94}},\ \bibinfo {pages} {116012} (\bibinfo {year} {2016})},\ \Eprint
  {http://arxiv.org/abs/1605.02379} {arXiv:1605.02379 [cond-mat.mes-hall]}
  \BibitemShut {NoStop}%
\bibitem [{\citenamefont {Akal}\ \emph {et~al.}(2017)\citenamefont {Akal},
  \citenamefont {Egger}, \citenamefont {M{\"u}ller},\ and\ \citenamefont
  {Villalba-Ch{\'a}vez}}]{Akal:2017vem}%
  \BibitemOpen
  \bibfield  {author} {\bibinfo {author} {\bibfnamefont {I.}~\bibnamefont
  {Akal}}, \bibinfo {author} {\bibfnamefont {R.}~\bibnamefont {Egger}},
  \bibinfo {author} {\bibfnamefont {C.}~\bibnamefont {M{\"u}ller}}, \ and\
  \bibinfo {author} {\bibfnamefont {S.}~\bibnamefont {Villalba-Ch{\'a}vez}},\
  }\href@noop {} {} (\bibinfo {year} {2017}),\ \bibinfo {note} {in
  preparation}\BibitemShut {NoStop}%
\bibitem [{\citenamefont {Oka}\ and\ \citenamefont
  {Aoki}(2005)}]{PhysRevLett.95.137601}%
  \BibitemOpen
  \bibfield  {author} {\bibinfo {author} {\bibfnamefont {T.}~\bibnamefont
  {Oka}}\ and\ \bibinfo {author} {\bibfnamefont {H.}~\bibnamefont {Aoki}},\
  }\href {\doibase 10.1103/PhysRevLett.95.137601} {\bibfield  {journal}
  {\bibinfo  {journal} {Phys. Rev. Lett.}\ }\textbf {\bibinfo {volume} {95}},\
  \bibinfo {pages} {137601} (\bibinfo {year} {2005})}\BibitemShut {NoStop}%
\bibitem [{\citenamefont {Sch{\"u}tzhold}\ \emph {et~al.}(2008)\citenamefont
  {Sch{\"u}tzhold}, \citenamefont {Gies},\ and\ \citenamefont
  {Dunne}}]{Schutzhold:2008pz}%
  \BibitemOpen
  \bibfield  {author} {\bibinfo {author} {\bibfnamefont {R.}~\bibnamefont
  {Sch{\"u}tzhold}}, \bibinfo {author} {\bibfnamefont {H.}~\bibnamefont
  {Gies}}, \ and\ \bibinfo {author} {\bibfnamefont {G.}~\bibnamefont {Dunne}},\
  }\href {\doibase 10.1103/PhysRevLett.101.130404} {\bibfield  {journal}
  {\bibinfo  {journal} {Phys. Rev. Lett.}\ }\textbf {\bibinfo {volume} {101}},\
  \bibinfo {pages} {130404} (\bibinfo {year} {2008})},\ \Eprint
  {http://arxiv.org/abs/0807.0754} {arXiv:0807.0754 [hep-th]} \BibitemShut
  {NoStop}%
\bibitem [{\citenamefont {Dunne}\ \emph {et~al.}(2009)\citenamefont {Dunne},
  \citenamefont {Gies},\ and\ \citenamefont {Sch{\"u}tzhold}}]{Dunne:2009gi}%
  \BibitemOpen
  \bibfield  {author} {\bibinfo {author} {\bibfnamefont {G.~V.}\ \bibnamefont
  {Dunne}}, \bibinfo {author} {\bibfnamefont {H.}~\bibnamefont {Gies}}, \ and\
  \bibinfo {author} {\bibfnamefont {R.}~\bibnamefont {Sch{\"u}tzhold}},\ }\href
  {\doibase 10.1103/PhysRevD.80.111301} {\bibfield  {journal} {\bibinfo
  {journal} {Phys. Rev.}\ }\textbf {\bibinfo {volume} {D80}},\ \bibinfo {pages}
  {111301} (\bibinfo {year} {2009})},\ \Eprint {http://arxiv.org/abs/0908.0948}
  {arXiv:0908.0948 [hep-ph]} \BibitemShut {NoStop}%
\bibitem [{\citenamefont {Bulanov}\ \emph {et~al.}(2010)\citenamefont
  {Bulanov}, \citenamefont {Mur}, \citenamefont {Narozhny}, \citenamefont
  {Nees},\ and\ \citenamefont {Popov}}]{Bulanov:2010ei}%
  \BibitemOpen
  \bibfield  {author} {\bibinfo {author} {\bibfnamefont {S.~S.}\ \bibnamefont
  {Bulanov}}, \bibinfo {author} {\bibfnamefont {V.~D.}\ \bibnamefont {Mur}},
  \bibinfo {author} {\bibfnamefont {N.~B.}\ \bibnamefont {Narozhny}}, \bibinfo
  {author} {\bibfnamefont {J.}~\bibnamefont {Nees}}, \ and\ \bibinfo {author}
  {\bibfnamefont {V.~S.}\ \bibnamefont {Popov}},\ }\href {\doibase
  10.1103/PhysRevLett.104.220404} {\bibfield  {journal} {\bibinfo  {journal}
  {Phys. Rev. Lett.}\ }\textbf {\bibinfo {volume} {104}},\ \bibinfo {pages}
  {220404} (\bibinfo {year} {2010})},\ \Eprint {http://arxiv.org/abs/1003.2623}
  {arXiv:1003.2623 [hep-ph]} \BibitemShut {NoStop}%
\bibitem [{\citenamefont {Akal}\ \emph {et~al.}(2014)\citenamefont {Akal},
  \citenamefont {Villalba-Ch{\'a}vez},\ and\ \citenamefont
  {M{\"u}ller}}]{Akal:2014eua}%
  \BibitemOpen
  \bibfield  {author} {\bibinfo {author} {\bibfnamefont {I.}~\bibnamefont
  {Akal}}, \bibinfo {author} {\bibfnamefont {S.}~\bibnamefont
  {Villalba-Ch{\'a}vez}}, \ and\ \bibinfo {author} {\bibfnamefont
  {C.}~\bibnamefont {M{\"u}ller}},\ }\href {\doibase
  10.1103/PhysRevD.90.113004} {\bibfield  {journal} {\bibinfo  {journal} {Phys.
  Rev.}\ }\textbf {\bibinfo {volume} {D90}},\ \bibinfo {pages} {113004}
  (\bibinfo {year} {2014})},\ \Eprint {http://arxiv.org/abs/1409.1806}
  {arXiv:1409.1806 [hep-ph]} \BibitemShut {NoStop}%
\bibitem [{\citenamefont {Otto}\ \emph {et~al.}(2015)\citenamefont {Otto},
  \citenamefont {Seipt}, \citenamefont {Blaschke}, \citenamefont {Smolyansky},\
  and\ \citenamefont {K{\"a}mpfer}}]{Otto:2015gla}%
  \BibitemOpen
  \bibfield  {author} {\bibinfo {author} {\bibfnamefont {A.}~\bibnamefont
  {Otto}}, \bibinfo {author} {\bibfnamefont {D.}~\bibnamefont {Seipt}},
  \bibinfo {author} {\bibfnamefont {D.}~\bibnamefont {Blaschke}}, \bibinfo
  {author} {\bibfnamefont {S.~A.}\ \bibnamefont {Smolyansky}}, \ and\ \bibinfo
  {author} {\bibfnamefont {B.}~\bibnamefont {K{\"a}mpfer}},\ }\href {\doibase
  10.1103/PhysRevD.91.105018} {\bibfield  {journal} {\bibinfo  {journal} {Phys.
  Rev.}\ }\textbf {\bibinfo {volume} {D91}},\ \bibinfo {pages} {105018}
  (\bibinfo {year} {2015})},\ \Eprint {http://arxiv.org/abs/1503.08675}
  {arXiv:1503.08675 [hep-ph]} \BibitemShut {NoStop}%
\bibitem [{\citenamefont {Linder}\ \emph {et~al.}(2015)\citenamefont {Linder},
  \citenamefont {Schneider}, \citenamefont {Sicking}, \citenamefont {Szpak},\
  and\ \citenamefont {Sch{\"u}tzhold}}]{Linder:2015vta}%
  \BibitemOpen
  \bibfield  {author} {\bibinfo {author} {\bibfnamefont {M.~F.}\ \bibnamefont
  {Linder}}, \bibinfo {author} {\bibfnamefont {C.}~\bibnamefont {Schneider}},
  \bibinfo {author} {\bibfnamefont {J.}~\bibnamefont {Sicking}}, \bibinfo
  {author} {\bibfnamefont {N.}~\bibnamefont {Szpak}}, \ and\ \bibinfo {author}
  {\bibfnamefont {R.}~\bibnamefont {Sch{\"u}tzhold}},\ }\href {\doibase
  10.1103/PhysRevD.92.085009} {\bibfield  {journal} {\bibinfo  {journal} {Phys.
  Rev.}\ }\textbf {\bibinfo {volume} {D92}},\ \bibinfo {pages} {085009}
  (\bibinfo {year} {2015})},\ \Eprint {http://arxiv.org/abs/1505.05685}
  {arXiv:1505.05685 [hep-th]} \BibitemShut {NoStop}%
\bibitem [{\citenamefont {Kohlf{\"u}rst}\ \emph {et~al.}(2013)\citenamefont
  {Kohlf{\"u}rst}, \citenamefont {Mitter}, \citenamefont {von Winckel},
  \citenamefont {Hebenstreit},\ and\ \citenamefont
  {Alkofer}}]{Kohlfurst:2012rb}%
  \BibitemOpen
  \bibfield  {author} {\bibinfo {author} {\bibfnamefont {C.}~\bibnamefont
  {Kohlf{\"u}rst}}, \bibinfo {author} {\bibfnamefont {M.}~\bibnamefont
  {Mitter}}, \bibinfo {author} {\bibfnamefont {G.}~\bibnamefont {von Winckel}},
  \bibinfo {author} {\bibfnamefont {F.}~\bibnamefont {Hebenstreit}}, \ and\
  \bibinfo {author} {\bibfnamefont {R.}~\bibnamefont {Alkofer}},\ }\href
  {\doibase 10.1103/PhysRevD.88.045028} {\bibfield  {journal} {\bibinfo
  {journal} {Phys. Rev.}\ }\textbf {\bibinfo {volume} {D88}},\ \bibinfo {pages}
  {045028} (\bibinfo {year} {2013})},\ \Eprint {http://arxiv.org/abs/1212.1385}
  {arXiv:1212.1385 [hep-ph]} \BibitemShut {NoStop}%
\bibitem [{\citenamefont {Hebenstreit}\ and\ \citenamefont
  {Fillion-Gourdeau}(2014)}]{Hebenstreit:2014lra}%
  \BibitemOpen
  \bibfield  {author} {\bibinfo {author} {\bibfnamefont {F.}~\bibnamefont
  {Hebenstreit}}\ and\ \bibinfo {author} {\bibfnamefont {F.}~\bibnamefont
  {Fillion-Gourdeau}},\ }\href {\doibase 10.1016/j.physletb.2014.10.056}
  {\bibfield  {journal} {\bibinfo  {journal} {Phys. Lett.}\ }\textbf {\bibinfo
  {volume} {B739}},\ \bibinfo {pages} {189} (\bibinfo {year} {2014})},\ \Eprint
  {http://arxiv.org/abs/1409.7943} {arXiv:1409.7943 [hep-ph]} \BibitemShut
  {NoStop}%
\bibitem [{\citenamefont {Fillion-Gourdeau}\ \emph {et~al.}(2017)\citenamefont
  {Fillion-Gourdeau}, \citenamefont {Hebenstreit}, \citenamefont {Gagnon},\
  and\ \citenamefont {MacLean}}]{Fillion-Gourdeau:2017uss}%
  \BibitemOpen
  \bibfield  {author} {\bibinfo {author} {\bibfnamefont {F.}~\bibnamefont
  {Fillion-Gourdeau}}, \bibinfo {author} {\bibfnamefont {F.}~\bibnamefont
  {Hebenstreit}}, \bibinfo {author} {\bibfnamefont {D.}~\bibnamefont {Gagnon}},
  \ and\ \bibinfo {author} {\bibfnamefont {S.}~\bibnamefont {MacLean}},\
  }\href@noop {} {\  (\bibinfo {year} {2017})},\ \Eprint
  {http://arxiv.org/abs/1704.08919} {arXiv:1704.08919 [hep-ph]} \BibitemShut
  {NoStop}%
\bibitem [{\citenamefont {Akal}\ and\ \citenamefont
  {Moortgat-Pick}(2017)}]{Akal:2017ilh}%
  \BibitemOpen
  \bibfield  {author} {\bibinfo {author} {\bibfnamefont {I.}~\bibnamefont
  {Akal}}\ and\ \bibinfo {author} {\bibfnamefont {G.}~\bibnamefont
  {Moortgat-Pick}},\ }\href@noop {} {\  (\bibinfo {year} {2017})},\ \Eprint
  {http://arxiv.org/abs/1706.06447} {arXiv:1706.06447 [hep-th]} \BibitemShut
  {NoStop}%
\bibitem [{\citenamefont {Kim}\ and\ \citenamefont {Page}(2002)}]{Kim:2000un}%
  \BibitemOpen
  \bibfield  {author} {\bibinfo {author} {\bibfnamefont {S.~P.}\ \bibnamefont
  {Kim}}\ and\ \bibinfo {author} {\bibfnamefont {D.~N.}\ \bibnamefont {Page}},\
  }\href {\doibase 10.1103/PhysRevD.65.105002} {\bibfield  {journal} {\bibinfo
  {journal} {Phys. Rev.}\ }\textbf {\bibinfo {volume} {D65}},\ \bibinfo {pages}
  {105002} (\bibinfo {year} {2002})},\ \Eprint
  {http://arxiv.org/abs/hep-th/0005078} {arXiv:hep-th/0005078 [hep-th]}
  \BibitemShut {NoStop}%
\bibitem [{\citenamefont {Kim}\ and\ \citenamefont {Page}(2006)}]{Kim:2003qp}%
  \BibitemOpen
  \bibfield  {author} {\bibinfo {author} {\bibfnamefont {S.~P.}\ \bibnamefont
  {Kim}}\ and\ \bibinfo {author} {\bibfnamefont {D.~N.}\ \bibnamefont {Page}},\
  }\href {\doibase 10.1103/PhysRevD.73.065020} {\bibfield  {journal} {\bibinfo
  {journal} {Phys. Rev.}\ }\textbf {\bibinfo {volume} {D73}},\ \bibinfo {pages}
  {065020} (\bibinfo {year} {2006})},\ \Eprint
  {http://arxiv.org/abs/hep-th/0301132} {arXiv:hep-th/0301132 [hep-th]}
  \BibitemShut {NoStop}%
\bibitem [{\citenamefont {Dunne}(2004)}]{Dunne:2004nc}%
  \BibitemOpen
  \bibfield  {author} {\bibinfo {author} {\bibfnamefont {G.~V.}\ \bibnamefont
  {Dunne}}\ }(\bibinfo {year} {2004})\ pp.\ \bibinfo {pages} {445--522},\
  \Eprint {http://arxiv.org/abs/hep-th/0406216} {arXiv:hep-th/0406216 [hep-th]}
  \BibitemShut {NoStop}%
\bibitem [{\citenamefont {Ruf}\ \emph {et~al.}(2009)\citenamefont {Ruf},
  \citenamefont {Mocken}, \citenamefont {M{\"u}ller}, \citenamefont
  {Hatsagortsyan},\ and\ \citenamefont {Keitel}}]{Ruf:2009zz}%
  \BibitemOpen
  \bibfield  {author} {\bibinfo {author} {\bibfnamefont {M.}~\bibnamefont
  {Ruf}}, \bibinfo {author} {\bibfnamefont {G.~R.}\ \bibnamefont {Mocken}},
  \bibinfo {author} {\bibfnamefont {C.}~\bibnamefont {M{\"u}ller}}, \bibinfo
  {author} {\bibfnamefont {K.~Z.}\ \bibnamefont {Hatsagortsyan}}, \ and\
  \bibinfo {author} {\bibfnamefont {C.~H.}\ \bibnamefont {Keitel}},\ }\href
  {\doibase 10.1103/PhysRevLett.102.080402} {\bibfield  {journal} {\bibinfo
  {journal} {Phys. Rev. Lett.}\ }\textbf {\bibinfo {volume} {102}},\ \bibinfo
  {pages} {080402} (\bibinfo {year} {2009})},\ \Eprint
  {http://arxiv.org/abs/0810.4047} {arXiv:0810.4047 [physics.atom-ph]}
  \BibitemShut {NoStop}%
\bibitem [{\citenamefont {Hebenstreit}\ \emph {et~al.}(2010)\citenamefont
  {Hebenstreit}, \citenamefont {Alkofer},\ and\ \citenamefont
  {Gies}}]{Hebenstreit:2010vz}%
  \BibitemOpen
  \bibfield  {author} {\bibinfo {author} {\bibfnamefont {F.}~\bibnamefont
  {Hebenstreit}}, \bibinfo {author} {\bibfnamefont {R.}~\bibnamefont
  {Alkofer}}, \ and\ \bibinfo {author} {\bibfnamefont {H.}~\bibnamefont
  {Gies}},\ }\href {\doibase 10.1103/PhysRevD.82.105026} {\bibfield  {journal}
  {\bibinfo  {journal} {Phys. Rev.}\ }\textbf {\bibinfo {volume} {D82}},\
  \bibinfo {pages} {105026} (\bibinfo {year} {2010})},\ \Eprint
  {http://arxiv.org/abs/1007.1099} {arXiv:1007.1099 [hep-ph]} \BibitemShut
  {NoStop}%
\bibitem [{\citenamefont {Hebenstreit}\ \emph {et~al.}(2011)\citenamefont
  {Hebenstreit}, \citenamefont {Alkofer},\ and\ \citenamefont
  {Gies}}]{Hebenstreit:2011wk}%
  \BibitemOpen
  \bibfield  {author} {\bibinfo {author} {\bibfnamefont {F.}~\bibnamefont
  {Hebenstreit}}, \bibinfo {author} {\bibfnamefont {R.}~\bibnamefont
  {Alkofer}}, \ and\ \bibinfo {author} {\bibfnamefont {H.}~\bibnamefont
  {Gies}},\ }\href {\doibase 10.1103/PhysRevLett.107.180403} {\bibfield
  {journal} {\bibinfo  {journal} {Phys. Rev. Lett.}\ }\textbf {\bibinfo
  {volume} {107}},\ \bibinfo {pages} {180403} (\bibinfo {year} {2011})},\
  \Eprint {http://arxiv.org/abs/1106.6175} {arXiv:1106.6175 [hep-ph]}
  \BibitemShut {NoStop}%
\bibitem [{\citenamefont {Brezin}\ and\ \citenamefont
  {Itzykson}(1970)}]{PhysRevD.2.1191}%
  \BibitemOpen
  \bibfield  {author} {\bibinfo {author} {\bibfnamefont {E.}~\bibnamefont
  {Brezin}}\ and\ \bibinfo {author} {\bibfnamefont {C.}~\bibnamefont
  {Itzykson}},\ }\href {\doibase 10.1103/PhysRevD.2.1191} {\bibfield  {journal}
  {\bibinfo  {journal} {Phys. Rev. D}\ }\textbf {\bibinfo {volume} {2}},\
  \bibinfo {pages} {1191} (\bibinfo {year} {1970})}\BibitemShut {NoStop}%
\bibitem [{\citenamefont {Popov}(1972)}]{Popov:1971iga}%
  \BibitemOpen
  \bibfield  {author} {\bibinfo {author} {\bibfnamefont {V.~S.}\ \bibnamefont
  {Popov}},\ }\href@noop {} {\bibfield  {journal} {\bibinfo  {journal} {Sov.
  Phys. JETP.}\ }\textbf {\bibinfo {volume} {34}},\ \bibinfo {pages} {709}
  (\bibinfo {year} {1972})},\ \bibinfo {note} {[Zh. Eksp. Teor.
  Fiz.61,1334(1971)]}\BibitemShut {NoStop}%
\bibitem [{\citenamefont {Marinov}\ and\ \citenamefont
  {Popov}(1977)}]{Marinov:1977gq}%
  \BibitemOpen
  \bibfield  {author} {\bibinfo {author} {\bibfnamefont {M.~S.}\ \bibnamefont
  {Marinov}}\ and\ \bibinfo {author} {\bibfnamefont {V.~S.}\ \bibnamefont
  {Popov}},\ }\href {\doibase 10.1002/prop.19770250111} {\bibfield  {journal}
  {\bibinfo  {journal} {Fortsch. Phys.}\ }\textbf {\bibinfo {volume} {25}},\
  \bibinfo {pages} {373} (\bibinfo {year} {1977})}\BibitemShut {NoStop}%
\bibitem [{\citenamefont {Keski-Vakkuri}\ and\ \citenamefont
  {Kraus}(1996)}]{KeskiVakkuri:1996gn}%
  \BibitemOpen
  \bibfield  {author} {\bibinfo {author} {\bibfnamefont {E.}~\bibnamefont
  {Keski-Vakkuri}}\ and\ \bibinfo {author} {\bibfnamefont {P.}~\bibnamefont
  {Kraus}},\ }\href {\doibase 10.1103/PhysRevD.54.7407} {\bibfield  {journal}
  {\bibinfo  {journal} {Phys. Rev.}\ }\textbf {\bibinfo {volume} {D54}},\
  \bibinfo {pages} {7407} (\bibinfo {year} {1996})},\ \Eprint
  {http://arxiv.org/abs/hep-th/9604151} {arXiv:hep-th/9604151 [hep-th]}
  \BibitemShut {NoStop}%
\bibitem [{\citenamefont {Bai-Song}\ \emph {et~al.}(2012)\citenamefont
  {Bai-Song}, \citenamefont {Melike},\ and\ \citenamefont
  {Sayipjamal}}]{bai2012electron}%
  \BibitemOpen
  \bibfield  {author} {\bibinfo {author} {\bibfnamefont {X.}~\bibnamefont
  {Bai-Song}}, \bibinfo {author} {\bibfnamefont {M.}~\bibnamefont {Melike}}, \
  and\ \bibinfo {author} {\bibfnamefont {D.}~\bibnamefont {Sayipjamal}},\
  }\href@noop {} {\bibfield  {journal} {\bibinfo  {journal} {Chinese Physics
  Letters}\ }\textbf {\bibinfo {volume} {29}},\ \bibinfo {pages} {021102}
  (\bibinfo {year} {2012})}\BibitemShut {NoStop}%
\bibitem [{\citenamefont {Dietrich}(2003)}]{Dietrich:2003qf}%
  \BibitemOpen
  \bibfield  {author} {\bibinfo {author} {\bibfnamefont {D.~D.}\ \bibnamefont
  {Dietrich}},\ }\href {\doibase 10.1103/PhysRevD.68.105005} {\bibfield
  {journal} {\bibinfo  {journal} {Phys. Rev.}\ }\textbf {\bibinfo {volume}
  {D68}},\ \bibinfo {pages} {105005} (\bibinfo {year} {2003})},\ \Eprint
  {http://arxiv.org/abs/hep-th/0302229} {arXiv:hep-th/0302229 [hep-th]}
  \BibitemShut {NoStop}%
\bibitem [{\citenamefont {Dunne}\ and\ \citenamefont
  {Schubert}(2005)}]{Dunne:2005sx}%
  \BibitemOpen
  \bibfield  {author} {\bibinfo {author} {\bibfnamefont {G.~V.}\ \bibnamefont
  {Dunne}}\ and\ \bibinfo {author} {\bibfnamefont {C.}~\bibnamefont
  {Schubert}},\ }\href {\doibase 10.1103/PhysRevD.72.105004} {\bibfield
  {journal} {\bibinfo  {journal} {Phys. Rev.}\ }\textbf {\bibinfo {volume}
  {D72}},\ \bibinfo {pages} {105004} (\bibinfo {year} {2005})},\ \Eprint
  {http://arxiv.org/abs/hep-th/0507174} {arXiv:hep-th/0507174 [hep-th]}
  \BibitemShut {NoStop}%
\bibitem [{\citenamefont {Kim}\ and\ \citenamefont
  {Schubert}(2011)}]{Kim:2011jw}%
  \BibitemOpen
  \bibfield  {author} {\bibinfo {author} {\bibfnamefont {S.~P.}\ \bibnamefont
  {Kim}}\ and\ \bibinfo {author} {\bibfnamefont {C.}~\bibnamefont {Schubert}},\
  }\href {\doibase 10.1103/PhysRevD.84.125028} {\bibfield  {journal} {\bibinfo
  {journal} {Phys. Rev.}\ }\textbf {\bibinfo {volume} {D84}},\ \bibinfo {pages}
  {125028} (\bibinfo {year} {2011})},\ \Eprint {http://arxiv.org/abs/1110.0900}
  {arXiv:1110.0900 [hep-th]} \BibitemShut {NoStop}%
\bibitem [{\citenamefont {Strobel}\ and\ \citenamefont
  {Xue}(2014)}]{Strobel:2013vza}%
  \BibitemOpen
  \bibfield  {author} {\bibinfo {author} {\bibfnamefont {E.}~\bibnamefont
  {Strobel}}\ and\ \bibinfo {author} {\bibfnamefont {S.-S.}\ \bibnamefont
  {Xue}},\ }\href {\doibase 10.1016/j.nuclphysb.2014.07.017} {\bibfield
  {journal} {\bibinfo  {journal} {Nucl. Phys.}\ }\textbf {\bibinfo {volume}
  {B886}},\ \bibinfo {pages} {1153} (\bibinfo {year} {2014})},\ \Eprint
  {http://arxiv.org/abs/1312.3261} {arXiv:1312.3261 [hep-th]} \BibitemShut
  {NoStop}%
\bibitem [{\citenamefont {Schneider}\ and\ \citenamefont
  {Sch{\"u}tzhold}(2016{\natexlab{a}})}]{Schneider:2014mla}%
  \BibitemOpen
  \bibfield  {author} {\bibinfo {author} {\bibfnamefont {C.}~\bibnamefont
  {Schneider}}\ and\ \bibinfo {author} {\bibfnamefont {R.}~\bibnamefont
  {Sch{\"u}tzhold}},\ }\href {\doibase 10.1007/JHEP02(2016)164} {\bibfield
  {journal} {\bibinfo  {journal} {JHEP}\ }\textbf {\bibinfo {volume} {02}},\
  \bibinfo {pages} {164} (\bibinfo {year} {2016}{\natexlab{a}})},\ \Eprint
  {http://arxiv.org/abs/1407.3584} {arXiv:1407.3584 [hep-th]} \BibitemShut
  {NoStop}%
\bibitem [{\citenamefont {Ilderton}\ \emph {et~al.}(2015)\citenamefont
  {Ilderton}, \citenamefont {Torgrimsson},\ and\ \citenamefont
  {Wardh}}]{Ilderton:2015qda}%
  \BibitemOpen
  \bibfield  {author} {\bibinfo {author} {\bibfnamefont {A.}~\bibnamefont
  {Ilderton}}, \bibinfo {author} {\bibfnamefont {G.}~\bibnamefont
  {Torgrimsson}}, \ and\ \bibinfo {author} {\bibfnamefont {J.}~\bibnamefont
  {Wardh}},\ }\href {\doibase 10.1103/PhysRevD.92.065001} {\bibfield  {journal}
  {\bibinfo  {journal} {Phys. Rev.}\ }\textbf {\bibinfo {volume} {D92}},\
  \bibinfo {pages} {065001} (\bibinfo {year} {2015})},\ \Eprint
  {http://arxiv.org/abs/1506.09186} {arXiv:1506.09186 [hep-th]} \BibitemShut
  {NoStop}%
\bibitem [{\citenamefont {Adorno}\ \emph {et~al.}(2016)\citenamefont {Adorno},
  \citenamefont {Gavrilov},\ and\ \citenamefont {Gitman}}]{Adorno:2016bjx}%
  \BibitemOpen
  \bibfield  {author} {\bibinfo {author} {\bibfnamefont {T.~C.}\ \bibnamefont
  {Adorno}}, \bibinfo {author} {\bibfnamefont {S.~P.}\ \bibnamefont
  {Gavrilov}}, \ and\ \bibinfo {author} {\bibfnamefont {D.~M.}\ \bibnamefont
  {Gitman}},\ }\href {\doibase 10.1140/epjc/s10052-016-4289-0} {\bibfield
  {journal} {\bibinfo  {journal} {Eur. Phys. J.}\ }\textbf {\bibinfo {volume}
  {C76}},\ \bibinfo {pages} {447} (\bibinfo {year} {2016})},\ \Eprint
  {http://arxiv.org/abs/1605.09072} {arXiv:1605.09072 [hep-th]} \BibitemShut
  {NoStop}%
\bibitem [{\citenamefont {Goriely}(2001)}]{goriely2001integrability}%
  \BibitemOpen
  \bibfield  {author} {\bibinfo {author} {\bibfnamefont {A.}~\bibnamefont
  {Goriely}},\ }\href@noop {} {\emph {\bibinfo {title} {Integrability and
  nonintegrability of dynamical systems}}},\ Vol.~\bibinfo {volume} {19}\
  (\bibinfo  {publisher} {World Scientific},\ \bibinfo {year}
  {2001})\BibitemShut {NoStop}%
\bibitem [{\citenamefont {Strassler}(1992)}]{Strassler:1992zr}%
  \BibitemOpen
  \bibfield  {author} {\bibinfo {author} {\bibfnamefont {M.~J.}\ \bibnamefont
  {Strassler}},\ }\href {\doibase 10.1016/0550-3213(92)90098-V} {\bibfield
  {journal} {\bibinfo  {journal} {Nucl. Phys.}\ }\textbf {\bibinfo {volume}
  {B385}},\ \bibinfo {pages} {145} (\bibinfo {year} {1992})},\ \Eprint
  {http://arxiv.org/abs/hep-ph/9205205} {arXiv:hep-ph/9205205 [hep-ph]}
  \BibitemShut {NoStop}%
\bibitem [{\citenamefont {Schubert}(2001)}]{Schubert:2001he}%
  \BibitemOpen
  \bibfield  {author} {\bibinfo {author} {\bibfnamefont {C.}~\bibnamefont
  {Schubert}},\ }\href {\doibase 10.1016/S0370-1573(01)00013-8} {\bibfield
  {journal} {\bibinfo  {journal} {Phys. Rept.}\ }\textbf {\bibinfo {volume}
  {355}},\ \bibinfo {pages} {73} (\bibinfo {year} {2001})},\ \Eprint
  {http://arxiv.org/abs/hep-th/0101036} {arXiv:hep-th/0101036 [hep-th]}
  \BibitemShut {NoStop}%
\bibitem [{\citenamefont {Affleck}\ \emph {et~al.}(1982)\citenamefont
  {Affleck}, \citenamefont {Alvarez},\ and\ \citenamefont
  {Manton}}]{Affleck:1981bma}%
  \BibitemOpen
  \bibfield  {author} {\bibinfo {author} {\bibfnamefont {I.~K.}\ \bibnamefont
  {Affleck}}, \bibinfo {author} {\bibfnamefont {O.}~\bibnamefont {Alvarez}}, \
  and\ \bibinfo {author} {\bibfnamefont {N.~S.}\ \bibnamefont {Manton}},\
  }\href {\doibase 10.1016/0550-3213(82)90455-2} {\bibfield  {journal}
  {\bibinfo  {journal} {Nucl. Phys.}\ }\textbf {\bibinfo {volume} {B197}},\
  \bibinfo {pages} {509} (\bibinfo {year} {1982})}\BibitemShut {NoStop}%
\bibitem [{\citenamefont {Dunne}\ \emph {et~al.}(2006)\citenamefont {Dunne},
  \citenamefont {Wang}, \citenamefont {Gies},\ and\ \citenamefont
  {Schubert}}]{Dunne:2006st}%
  \BibitemOpen
  \bibfield  {author} {\bibinfo {author} {\bibfnamefont {G.~V.}\ \bibnamefont
  {Dunne}}, \bibinfo {author} {\bibfnamefont {Q.-h.}\ \bibnamefont {Wang}},
  \bibinfo {author} {\bibfnamefont {H.}~\bibnamefont {Gies}}, \ and\ \bibinfo
  {author} {\bibfnamefont {C.}~\bibnamefont {Schubert}},\ }\href {\doibase
  10.1103/PhysRevD.73.065028} {\bibfield  {journal} {\bibinfo  {journal} {Phys.
  Rev.}\ }\textbf {\bibinfo {volume} {D73}},\ \bibinfo {pages} {065028}
  (\bibinfo {year} {2006})},\ \Eprint {http://arxiv.org/abs/hep-th/0602176}
  {arXiv:hep-th/0602176 [hep-th]} \BibitemShut {NoStop}%
\bibitem [{\citenamefont {Dumlu}\ and\ \citenamefont
  {Dunne}(2011{\natexlab{a}})}]{Dumlu:2011cc}%
  \BibitemOpen
  \bibfield  {author} {\bibinfo {author} {\bibfnamefont {C.~K.}\ \bibnamefont
  {Dumlu}}\ and\ \bibinfo {author} {\bibfnamefont {G.~V.}\ \bibnamefont
  {Dunne}},\ }\href {\doibase 10.1103/PhysRevD.84.125023} {\bibfield  {journal}
  {\bibinfo  {journal} {Phys. Rev.}\ }\textbf {\bibinfo {volume} {D84}},\
  \bibinfo {pages} {125023} (\bibinfo {year} {2011}{\natexlab{a}})},\ \Eprint
  {http://arxiv.org/abs/1110.1657} {arXiv:1110.1657 [hep-th]} \BibitemShut
  {NoStop}%
\bibitem [{\citenamefont {Ilderton}(2014)}]{Ilderton:2014mla}%
  \BibitemOpen
  \bibfield  {author} {\bibinfo {author} {\bibfnamefont {A.}~\bibnamefont
  {Ilderton}},\ }\href {\doibase 10.1007/JHEP09(2014)166} {\bibfield  {journal}
  {\bibinfo  {journal} {JHEP}\ }\textbf {\bibinfo {volume} {09}},\ \bibinfo
  {pages} {166} (\bibinfo {year} {2014})},\ \Eprint
  {http://arxiv.org/abs/1406.1513} {arXiv:1406.1513 [hep-th]} \BibitemShut
  {NoStop}%
\bibitem [{\citenamefont {Dietrich}(2014)}]{Dietrich:2014ala}%
  \BibitemOpen
  \bibfield  {author} {\bibinfo {author} {\bibfnamefont {D.~D.}\ \bibnamefont
  {Dietrich}},\ }\href {\doibase 10.1103/PhysRevD.90.045024} {\bibfield
  {journal} {\bibinfo  {journal} {Phys. Rev.}\ }\textbf {\bibinfo {volume}
  {D90}},\ \bibinfo {pages} {045024} (\bibinfo {year} {2014})},\ \Eprint
  {http://arxiv.org/abs/1405.0487} {arXiv:1405.0487 [hep-ph]} \BibitemShut
  {NoStop}%
\bibitem [{\citenamefont {Ba{\c s}ar}\ and\ \citenamefont
  {Dunne}(2015)}]{Basar:2015xna}%
  \BibitemOpen
  \bibfield  {author} {\bibinfo {author} {\bibfnamefont {G.}~\bibnamefont
  {Ba{\c s}ar}}\ and\ \bibinfo {author} {\bibfnamefont {G.~V.}\ \bibnamefont
  {Dunne}},\ }\href {\doibase 10.1007/JHEP02(2015)160} {\bibfield  {journal}
  {\bibinfo  {journal} {JHEP}\ }\textbf {\bibinfo {volume} {02}},\ \bibinfo
  {pages} {160} (\bibinfo {year} {2015})},\ \Eprint
  {http://arxiv.org/abs/1501.05671} {arXiv:1501.05671 [hep-th]} \BibitemShut
  {NoStop}%
\bibitem [{\citenamefont {Gutzwiller}(1971)}]{Gutzwiller:1971fy}%
  \BibitemOpen
  \bibfield  {author} {\bibinfo {author} {\bibfnamefont {M.~C.}\ \bibnamefont
  {Gutzwiller}},\ }\href {\doibase 10.1063/1.1665596} {\bibfield  {journal}
  {\bibinfo  {journal} {J. Math. Phys.}\ }\textbf {\bibinfo {volume} {12}},\
  \bibinfo {pages} {343} (\bibinfo {year} {1971})}\BibitemShut {NoStop}%
\bibitem [{\citenamefont {{Littlejohn}}(1990)}]{1990JMP31.2952L}%
  \BibitemOpen
  \bibfield  {author} {\bibinfo {author} {\bibfnamefont {R.~G.}\ \bibnamefont
  {{Littlejohn}}},\ }\href {\doibase 10.1063/1.528949} {\bibfield  {journal}
  {\bibinfo  {journal} {Journal of Mathematical Physics}\ }\textbf {\bibinfo
  {volume} {31}},\ \bibinfo {pages} {2952} (\bibinfo {year}
  {1990})}\BibitemShut {NoStop}%
\bibitem [{\citenamefont {Aurich}\ and\ \citenamefont
  {Steiner}(1988)}]{Aurich:1988yv}%
  \BibitemOpen
  \bibfield  {author} {\bibinfo {author} {\bibfnamefont {R.}~\bibnamefont
  {Aurich}}\ and\ \bibinfo {author} {\bibfnamefont {F.}~\bibnamefont
  {Steiner}},\ }\href@noop {} {\bibfield  {journal} {\bibinfo  {journal}
  {Physica}\ }\textbf {\bibinfo {volume} {D32}},\ \bibinfo {pages} {451}
  (\bibinfo {year} {1988})}\BibitemShut {NoStop}%
\bibitem [{\citenamefont {Sieber}\ and\ \citenamefont
  {Steiner}(1990)}]{SIEBER1990159}%
  \BibitemOpen
  \bibfield  {author} {\bibinfo {author} {\bibfnamefont {M.}~\bibnamefont
  {Sieber}}\ and\ \bibinfo {author} {\bibfnamefont {F.}~\bibnamefont
  {Steiner}},\ }\href {\doibase http://dx.doi.org/10.1016/0375-9601(90)90692-H}
  {\bibfield  {journal} {\bibinfo  {journal} {Physics Letters A}\ }\textbf
  {\bibinfo {volume} {144}},\ \bibinfo {pages} {159 } (\bibinfo {year}
  {1990})}\BibitemShut {NoStop}%
\bibitem [{\citenamefont {Richens}\ and\ \citenamefont
  {Berry}(1981)}]{RICHENS1981495}%
  \BibitemOpen
  \bibfield  {author} {\bibinfo {author} {\bibfnamefont {P.}~\bibnamefont
  {Richens}}\ and\ \bibinfo {author} {\bibfnamefont {M.}~\bibnamefont
  {Berry}},\ }\href {\doibase http://dx.doi.org/10.1016/0167-2789(81)90024-5}
  {\bibfield  {journal} {\bibinfo  {journal} {Physica D: Nonlinear Phenomena}\
  }\textbf {\bibinfo {volume} {2}},\ \bibinfo {pages} {495 } (\bibinfo {year}
  {1981})}\BibitemShut {NoStop}%
\bibitem [{\citenamefont {M{\"u}ller}\ \emph {et~al.}(2004)\citenamefont
  {M{\"u}ller}, \citenamefont {Heusler}, \citenamefont {Braun}, \citenamefont
  {Haake},\ and\ \citenamefont {Altland}}]{muller2004semiclassical}%
  \BibitemOpen
  \bibfield  {author} {\bibinfo {author} {\bibfnamefont {S.}~\bibnamefont
  {M{\"u}ller}}, \bibinfo {author} {\bibfnamefont {S.}~\bibnamefont {Heusler}},
  \bibinfo {author} {\bibfnamefont {P.}~\bibnamefont {Braun}}, \bibinfo
  {author} {\bibfnamefont {F.}~\bibnamefont {Haake}}, \ and\ \bibinfo {author}
  {\bibfnamefont {A.}~\bibnamefont {Altland}},\ }\href@noop {} {\bibfield
  {journal} {\bibinfo  {journal} {Physical review letters}\ }\textbf {\bibinfo
  {volume} {93}},\ \bibinfo {pages} {014103} (\bibinfo {year}
  {2004})}\BibitemShut {NoStop}%
\bibitem [{\citenamefont {M{\"u}ller}\ \emph {et~al.}(2009)\citenamefont
  {M{\"u}ller}, \citenamefont {Heusler}, \citenamefont {Altland}, \citenamefont
  {Braun},\ and\ \citenamefont {Haake}}]{muller2009periodic}%
  \BibitemOpen
  \bibfield  {author} {\bibinfo {author} {\bibfnamefont {S.}~\bibnamefont
  {M{\"u}ller}}, \bibinfo {author} {\bibfnamefont {S.}~\bibnamefont {Heusler}},
  \bibinfo {author} {\bibfnamefont {A.}~\bibnamefont {Altland}}, \bibinfo
  {author} {\bibfnamefont {P.}~\bibnamefont {Braun}}, \ and\ \bibinfo {author}
  {\bibfnamefont {F.}~\bibnamefont {Haake}},\ }\href@noop {} {\bibfield
  {journal} {\bibinfo  {journal} {New Journal of Physics}\ }\textbf {\bibinfo
  {volume} {11}},\ \bibinfo {pages} {103025} (\bibinfo {year}
  {2009})}\BibitemShut {NoStop}%
\bibitem [{\citenamefont {Dietrich}\ and\ \citenamefont
  {Dunne}(2007)}]{Dietrich:2007vw}%
  \BibitemOpen
  \bibfield  {author} {\bibinfo {author} {\bibfnamefont {D.~D.}\ \bibnamefont
  {Dietrich}}\ and\ \bibinfo {author} {\bibfnamefont {G.~V.}\ \bibnamefont
  {Dunne}},\ }\href {\doibase 10.1088/1751-8113/40/34/F01} {\bibfield
  {journal} {\bibinfo  {journal} {J. Phys.}\ }\textbf {\bibinfo {volume}
  {A40}},\ \bibinfo {pages} {F825} (\bibinfo {year} {2007})},\ \Eprint
  {http://arxiv.org/abs/0706.4006} {arXiv:0706.4006 [hep-th]} \BibitemShut
  {NoStop}%
\bibitem [{\citenamefont {Berry}\ and\ \citenamefont
  {Tabor}(1976)}]{berry1976closed}%
  \BibitemOpen
  \bibfield  {author} {\bibinfo {author} {\bibfnamefont {M.~V.}\ \bibnamefont
  {Berry}}\ and\ \bibinfo {author} {\bibfnamefont {M.}~\bibnamefont {Tabor}},\
  }\bibfield  {booktitle} {\emph {\bibinfo {booktitle} {Proceedings of the
  Royal Society of London A: Mathematical, Physical and Engineering
  Sciences}},\ }\href@noop {} {\ \textbf {\bibinfo {volume} {349}},\ \bibinfo
  {pages} {101} (\bibinfo {year} {1976})}\BibitemShut {NoStop}%
\bibitem [{\citenamefont
  {Muratore-Ginanneschi}(2003)}]{MuratoreGinanneschi:2002tm}%
  \BibitemOpen
  \bibfield  {author} {\bibinfo {author} {\bibfnamefont {P.}~\bibnamefont
  {Muratore-Ginanneschi}},\ }\href {\doibase 10.1016/S0370-1573(03)00212-6}
  {\bibfield  {journal} {\bibinfo  {journal} {Phys. Rept.}\ }\textbf {\bibinfo
  {volume} {383}},\ \bibinfo {pages} {299} (\bibinfo {year} {2003})},\ \Eprint
  {http://arxiv.org/abs/nlin/0210047} {arXiv:nlin/0210047 [nlin-cd]}
  \BibitemShut {NoStop}%
\bibitem [{\citenamefont {Dumlu}(2016)}]{Dumlu:2015paa}%
  \BibitemOpen
  \bibfield  {author} {\bibinfo {author} {\bibfnamefont {C.~K.}\ \bibnamefont
  {Dumlu}},\ }\href {\doibase 10.1103/PhysRevD.93.065045} {\bibfield  {journal}
  {\bibinfo  {journal} {Phys. Rev.}\ }\textbf {\bibinfo {volume} {D93}},\
  \bibinfo {pages} {065045} (\bibinfo {year} {2016})},\ \Eprint
  {http://arxiv.org/abs/1507.07005} {arXiv:1507.07005 [hep-th]} \BibitemShut
  {NoStop}%
\bibitem [{\citenamefont {Shivamoggi}(2014)}]{shivamoggi2014nonlinear}%
  \BibitemOpen
  \bibfield  {author} {\bibinfo {author} {\bibfnamefont {B.~K.}\ \bibnamefont
  {Shivamoggi}},\ }\href@noop {} {\emph {\bibinfo {title} {Nonlinear dynamics
  and chaotic phenomena: An introduction}}},\ Vol.\ \bibinfo {volume} {103}\
  (\bibinfo  {publisher} {Springer},\ \bibinfo {year} {2014})\BibitemShut
  {NoStop}%
\bibitem [{\citenamefont {Sun}(2017)}]{Sun2017}%
  \BibitemOpen
  \bibfield  {author} {\bibinfo {author} {\bibfnamefont {S.}~\bibnamefont
  {Sun}},\ }\href {\doibase 10.1007/s11784-016-0355-3} {\bibfield  {journal}
  {\bibinfo  {journal} {Journal of Fixed Point Theory and Applications}\
  }\textbf {\bibinfo {volume} {19}},\ \bibinfo {pages} {299} (\bibinfo {year}
  {2017})}\BibitemShut {NoStop}%
\bibitem [{\citenamefont {Cohen}\ and\ \citenamefont
  {McGady}(2008)}]{Cohen:2008wz}%
  \BibitemOpen
  \bibfield  {author} {\bibinfo {author} {\bibfnamefont {T.~D.}\ \bibnamefont
  {Cohen}}\ and\ \bibinfo {author} {\bibfnamefont {D.~A.}\ \bibnamefont
  {McGady}},\ }\href {\doibase 10.1103/PhysRevD.78.036008} {\bibfield
  {journal} {\bibinfo  {journal} {Phys. Rev.}\ }\textbf {\bibinfo {volume}
  {D78}},\ \bibinfo {pages} {036008} (\bibinfo {year} {2008})},\ \Eprint
  {http://arxiv.org/abs/0807.1117} {arXiv:0807.1117 [hep-ph]} \BibitemShut
  {NoStop}%
\bibitem [{\citenamefont {Schneider}\ and\ \citenamefont
  {Sch{\"u}tzhold}(2016{\natexlab{b}})}]{Schneider:2016vrl}%
  \BibitemOpen
  \bibfield  {author} {\bibinfo {author} {\bibfnamefont {C.}~\bibnamefont
  {Schneider}}\ and\ \bibinfo {author} {\bibfnamefont {R.}~\bibnamefont
  {Sch{\"u}tzhold}},\ }\href {\doibase 10.1103/PhysRevD.94.085015} {\bibfield
  {journal} {\bibinfo  {journal} {Phys. Rev.}\ }\textbf {\bibinfo {volume}
  {D94}},\ \bibinfo {pages} {085015} (\bibinfo {year} {2016}{\natexlab{b}})},\
  \Eprint {http://arxiv.org/abs/1603.00864} {arXiv:1603.00864 [hep-th]}
  \BibitemShut {NoStop}%
\bibitem [{\citenamefont {Dumlu}\ and\ \citenamefont
  {Dunne}(2011{\natexlab{b}})}]{Dumlu:2011rr}%
  \BibitemOpen
  \bibfield  {author} {\bibinfo {author} {\bibfnamefont {C.~K.}\ \bibnamefont
  {Dumlu}}\ and\ \bibinfo {author} {\bibfnamefont {G.~V.}\ \bibnamefont
  {Dunne}},\ }\href {\doibase 10.1103/PhysRevD.83.065028} {\bibfield  {journal}
  {\bibinfo  {journal} {Phys. Rev.}\ }\textbf {\bibinfo {volume} {D83}},\
  \bibinfo {pages} {065028} (\bibinfo {year} {2011}{\natexlab{b}})},\ \Eprint
  {http://arxiv.org/abs/1102.2899} {arXiv:1102.2899 [hep-th]} \BibitemShut
  {NoStop}%
\bibitem [{\citenamefont {Orthaber}\ \emph {et~al.}(2011)\citenamefont
  {Orthaber}, \citenamefont {Hebenstreit},\ and\ \citenamefont
  {Alkofer}}]{Orthaber:2011cm}%
  \BibitemOpen
  \bibfield  {author} {\bibinfo {author} {\bibfnamefont {M.}~\bibnamefont
  {Orthaber}}, \bibinfo {author} {\bibfnamefont {F.}~\bibnamefont
  {Hebenstreit}}, \ and\ \bibinfo {author} {\bibfnamefont {R.}~\bibnamefont
  {Alkofer}},\ }\href {\doibase 10.1016/j.physletb.2011.02.053} {\bibfield
  {journal} {\bibinfo  {journal} {Phys. Lett.}\ }\textbf {\bibinfo {volume}
  {B698}},\ \bibinfo {pages} {80} (\bibinfo {year} {2011})},\ \Eprint
  {http://arxiv.org/abs/1102.2182} {arXiv:1102.2182 [hep-ph]} \BibitemShut
  {NoStop}%
\bibitem [{\citenamefont {Gies}\ and\ \citenamefont
  {Klingm{\"u}ller}(2005)}]{Gies:2005bz}%
  \BibitemOpen
  \bibfield  {author} {\bibinfo {author} {\bibfnamefont {H.}~\bibnamefont
  {Gies}}\ and\ \bibinfo {author} {\bibfnamefont {K.}~\bibnamefont
  {Klingm{\"u}ller}},\ }\href {\doibase 10.1103/PhysRevD.72.065001} {\bibfield
  {journal} {\bibinfo  {journal} {Phys. Rev.}\ }\textbf {\bibinfo {volume}
  {D72}},\ \bibinfo {pages} {065001} (\bibinfo {year} {2005})},\ \Eprint
  {http://arxiv.org/abs/hep-ph/0505099} {arXiv:hep-ph/0505099 [hep-ph]}
  \BibitemShut {NoStop}%
\bibitem [{\citenamefont {Nikishov}(1970)}]{NIKISHOV1970346}%
  \BibitemOpen
  \bibfield  {author} {\bibinfo {author} {\bibfnamefont {A.}~\bibnamefont
  {Nikishov}},\ }\href {\doibase
  http://dx.doi.org/10.1016/0550-3213(70)90527-4} {\bibfield  {journal}
  {\bibinfo  {journal} {Nuclear Physics B}\ }\textbf {\bibinfo {volume} {21}},\
  \bibinfo {pages} {346 } (\bibinfo {year} {1970})}\BibitemShut {NoStop}%
\bibitem [{\citenamefont {Gies}\ and\ \citenamefont
  {Torgrimsson}(2016)}]{Gies:2015hia}%
  \BibitemOpen
  \bibfield  {author} {\bibinfo {author} {\bibfnamefont {H.}~\bibnamefont
  {Gies}}\ and\ \bibinfo {author} {\bibfnamefont {G.}~\bibnamefont
  {Torgrimsson}},\ }\href {\doibase 10.1103/PhysRevLett.116.090406} {\bibfield
  {journal} {\bibinfo  {journal} {Phys. Rev. Lett.}\ }\textbf {\bibinfo
  {volume} {116}},\ \bibinfo {pages} {090406} (\bibinfo {year} {2016})},\
  \Eprint {http://arxiv.org/abs/1507.07802} {arXiv:1507.07802 [hep-ph]}
  \BibitemShut {NoStop}%
\bibitem [{\citenamefont {Tomaras}\ \emph {et~al.}(2001)\citenamefont
  {Tomaras}, \citenamefont {Tsamis},\ and\ \citenamefont
  {Woodard}}]{Tomaras:2001vs}%
  \BibitemOpen
  \bibfield  {author} {\bibinfo {author} {\bibfnamefont {T.~N.}\ \bibnamefont
  {Tomaras}}, \bibinfo {author} {\bibfnamefont {N.~C.}\ \bibnamefont {Tsamis}},
  \ and\ \bibinfo {author} {\bibfnamefont {R.~P.}\ \bibnamefont {Woodard}},\
  }\href {\doibase 10.1088/1126-6708/2001/11/008} {\bibfield  {journal}
  {\bibinfo  {journal} {JHEP}\ }\textbf {\bibinfo {volume} {11}},\ \bibinfo
  {pages} {008} (\bibinfo {year} {2001})},\ \Eprint
  {http://arxiv.org/abs/hep-th/0108090} {arXiv:hep-th/0108090 [hep-th]}
  \BibitemShut {NoStop}%
\bibitem [{\citenamefont {Dunne}\ and\ \citenamefont
  {Wang}(2006)}]{Dunne:2006ur}%
  \BibitemOpen
  \bibfield  {author} {\bibinfo {author} {\bibfnamefont {G.~V.}\ \bibnamefont
  {Dunne}}\ and\ \bibinfo {author} {\bibfnamefont {Q.-h.}\ \bibnamefont
  {Wang}},\ }\href {\doibase 10.1103/PhysRevD.74.065015} {\bibfield  {journal}
  {\bibinfo  {journal} {Phys. Rev.}\ }\textbf {\bibinfo {volume} {D74}},\
  \bibinfo {pages} {065015} (\bibinfo {year} {2006})},\ \Eprint
  {http://arxiv.org/abs/hep-th/0608020} {arXiv:hep-th/0608020 [hep-th]}
  \BibitemShut {NoStop}%
\bibitem [{\citenamefont {Torgrimsson}\ \emph {et~al.}(2017)\citenamefont
  {Torgrimsson}, \citenamefont {Schneider}, \citenamefont {Oertel},\ and\
  \citenamefont {Sch{\"u}tzhold}}]{Torgrimsson:2017pzs}%
  \BibitemOpen
  \bibfield  {author} {\bibinfo {author} {\bibfnamefont {G.}~\bibnamefont
  {Torgrimsson}}, \bibinfo {author} {\bibfnamefont {C.}~\bibnamefont
  {Schneider}}, \bibinfo {author} {\bibfnamefont {J.}~\bibnamefont {Oertel}}, \
  and\ \bibinfo {author} {\bibfnamefont {R.}~\bibnamefont {Sch{\"u}tzhold}},\
  }\href {\doibase 10.1007/JHEP06(2017)043} {\bibfield  {journal} {\bibinfo
  {journal} {JHEP}\ }\textbf {\bibinfo {volume} {06}},\ \bibinfo {pages} {043}
  (\bibinfo {year} {2017})},\ \Eprint {http://arxiv.org/abs/1703.09203}
  {arXiv:1703.09203 [hep-th]} \BibitemShut {NoStop}%
\bibitem [{\citenamefont {Akal}()}]{Akal-sg:2017}%
  \BibitemOpen
  \bibfield  {author} {\bibinfo {author} {\bibfnamefont {I.}~\bibnamefont
  {Akal}},\ }\href@noop {} {}\bibinfo {note} {(2017) in
  preparation}\BibitemShut {NoStop}%
\end{thebibliography}%
\end{document}